# Exchange Bias and Interface-related Effects in Two-dimensional van der Waals Magnetic Heterostructures: Open Questions and Perspectives

Manh-Huong Phan*, Vijaysankar Kalappattil, Valery Ortiz Jimenez, Yen Thi Hai Pham, Nivarthana W.Y.A.Y. Mudiyanselage, Derick Detellem, Chang-Ming Hung, Amit Chanda, and Tatiana Eggers

Department of Physics, University of South Florida, Tampa, Florida 33620, USA

The exchange bias (EB) effect is known as a fundamentally and technologically important magnetic property of a magnetic bilayer film. It is manifested as a horizontal shift in a magnetic hysteresis loop of a film subject to cooling in the presence of a magnetic field. The EB effect in van der Waals (vdW) heterostructures offers a novel approach for tuning the magnetic properties of the newly discovered single-layer magnets, as well as adds a new impetus to magnetic vdW heterostructures. Indeed, intriguing EB effects have recently been reported in a variety of low-dimensional vdW magnetic systems ranging from a weakly interlayer-coupled vdW magnet (e.g., $Fe_3GeTe_2$) to a bilayer composed of two different magnetic vdW materials (e.g., $Fe_3GeTe_2/CrCl_3$, $Fe_3GeTe_2/FePS_3$, $Fe_3GeTe_2/MnPS_3$), to bilayers of two different vdW defective magnets (e.g., $VSe_2/MoS_2$), or to metallic ferromagnet/vdW defective magnet interfaces (e.g., $Fe/MoS_2$). Despite their huge potential in spintronic device applications, the physical origins of the observed EB effects have remained elusive to researchers. We present here a critical review of the EB effect and associated phenomena such as magnetic proximity (MP) in various vdW heterostructure systems and propose approaches to addressing some of the emerging fundamental questions.

*Corresponding author: phanm@usf.edu

Table of Contents

**1. Introduction**

**2. Exchange Bias and Related Phenomena**

2.1. Weakly Interlayer-Coupled vdW Ferromagnets

2.2. Ferromagnetic vdW/Antiferromagnetic vdW Interfaces

2.3. Ferromagnetic vdW/Antiferromagnetic non-vdW Interfaces

2.4. Ferromagnetic vdW/Ferromagnetic non-vdW Interfaces

2.5. Ferromagnetic vdW/Defective Magnetic vdW Interfaces

2.6. Ferromagnetic non-vdW/Defective Magnetic vdW Interfaces

2.7. Defective Magnetic vdW Interfaces

**3. Perspective Applications**

**4. Concluding Remarks and Outlooks**

**References**

## 1. Introduction

Understanding interfacial magnetism in magnetic heterostructures is the key to unlocking the door to novel applications in spintronics and magnetic devices [1][2][3][4][5][6]. In traditional heterostructure materials, interface properties are often strongly modified by lattice strain, intermixing, or charge transfer, so that the magnetic properties at the interface are generally different from that of the bulk [1][7][8]. In these heterostructures, the intriguing exchange bias (EB) effect, often viewed as a horizontal shift in the magnetic hysteresis loop subject to cooling of a magnetic bilayer film in the presence of a magnetic field, has been reported [9][10][11][12][13][14][15]. Understanding the EB mechanism enables the optimization and manipulation of the magnetic functionality of spintronic devices, but this goal remains challenging due to uncontrolled interface properties or poor interface quality of these systems [13][14][15]. In this context, exploiting the EB effect in van der Waals (vdW) magnetic heterostructures is of paramount importance [16][17][18] as the high-quality interfaces and weakly coupled interlayer interactions in these systems allow probing intrinsically interfacial magnetic coupling mechanisms [19][20][21][22] that govern EB behavior (Fig. 1).

The EB effect has been known since 1956 when Meiklejohn and Bean observed a horizontal shift in the magnetic hysteresis loop of ferromagnetic/antiferromagnetic (FM/AFM) Co/CoO core/shell nanoparticles cooled in the presence of an external magnetic field [23]. Since, EB has been extensively studied in several FM/AFM thin films and core/shell systems, and its practical implementation has been realized in commercial magnetic recording and other spintronic devices [9][10][11]. For instance, EB has proved its usefulness in pinning a magnetically hard reference layer in spin valve (SV) readback heads and MRAM memory circuits, as well as improving the overall thermal stability of small magnetic particles within magnetic disk media through their FM/AFM interfacial coupling [24][25][26][27]. Despite these realized applications, the EB mechanism remains highly controversial [11][13]. In FM/AFM systems, the EB effect was initially ascribed to a unidirectional pinning of the FM layer by the adjacent AFM layer causing a shift in the switching field of the FM layer away from the $H = 0$ axis by an amount $H_{EB}$ (the EB field) and is accompanied by an increase in the coercive field ($H_C$) of the FM layer [9]. However, recent research has shown that the pinned uncompensated moments in FM/AFM systems generated by defects at the FM/AFM interface play a dominant role in determining the EB magnitude [28][29]. These disordered interfacial spins behave like "*spin clusters*" (SCs), which are analogous to spin glasses (SGs) at the FM/AFM interface. These SCs transmit the anisotropy from the AFM layer to the FM layer allowing for interactions via exchange coupling to the AFM and FM layers that gives rise to the coercivity of the FM layer [30][11]. Indeed, Berkowitz *et al.* found magnetically hard particles (e.g., $CoFe_2O_4$) at the FM/AFM interface in FeNi/CoO FM/AFM bilayers and attributed the EB effect and coercivity enhancement of the FM FeNi layer to the exchange coupling between the interfacial spins produced by the $CoFe_2O_4$ nanoparticles and those of the AFM (CoO) layer [31]. The authors also demonstrated that the interfacial spins produced by the FeNi layer did not contribute to the EB effect. In a recent review, Schuller *et al.* also revealed that the inner, uncompensated, pinned spins in the AFM layer play a crucial role in setting the EB magnitude in FM/AFM systems [29]. These arguments are supported by the experimental observations of the intrinsic EB effects in FM/SG interfaces (e.g., Co/CuMn [32] and $La_{0.7}Sr_{0.3}MnO_3$/LaNiO$_3$ [33]), and most recently in AFM/SG interfaces such as $Fe_xNbS_2$ [34]. In the case of $Fe_xNbS_2$, the EB has been shown to set in the SG phase that arose in the disordered AFM layer, and both the AFM and SG parameters could be tuned to manipulate the EB magnitude [34]. In a similar perspective, the coexistence of ferromagnetism and geometric frustration intrinsic to the Kagome network of magnetic ions has recently been demonstrated to give rise to SG behavior and the EB effect in the ferromagnetic Weyl semimetal $Co_3Sn_2S_2$ [35]. Geometric spin frustration has also been shown to give rise to a large EB effect in the conductive, layered 2D metal-organic Kagome lattice compound $Mn_3(C_6S_6)$ – which represents the first example of a topological SG showing EB [36]. These studies demonstrate that an AFM/FM interface is not the only condition that induces the EB effect and, in fact, EB is a more general phenomenon as other types of interfaces including ferromagnet/ferrimagnet FM/FIM, soft FM/hard FM, FIM/AFM, FIM/FIM, and AFM/dilute magnetic semiconductors (DMSs) have been reported to show EB characteristics as well [13][14]. To complicate matters, other effects, such as magnetic proximity (MP) [37][38][39][40][41][42] where magnetization from a ferromagnet penetrates a neighboring non-magnetic material or affects the magnetic ordering of an adjacent magnetic material, that have been reported to impact EB in magnetically interfaced systems [43][44] are often ignored.

The recent discoveries of two-dimensional (2D) intrinsic magnetism in atomically thin layers of $CrI_3$ [45] and $Cr_2Ge_2Te_6$ [46] have triggered global attention due to their prospective applications in spin transistors, valleytronics, and quantum computing [3][4][5][6][47][48]. Subsequently, a wide range of 2D magnetic materials (with both intrinsic and extrinsic magnetism) have been realized in monolayers of $VSe_2$ [49], $MnSe_2$ [50], $Cr_3Te_4$ [51], $Fe_3GeTe_2$ [52], $Fe_5GeTe_2$ [53], and $FePS_3$ [54]. Most recently, tunable room-temperature ferromagnetism has been discovered in magnetically-doped transition metal chalcogenide (TMD) monolayers including chemical vapor deposition- (CVD-) grown monolayers of V-doped $WSe_2$ [55][56], V-doped $WS_2$ [57], and Fe-doped $MoS_2$ [58]. These TMD monolayers are also known as 2D dilute magnetic semiconductors and they have potential to revolutionize the fields of spintronics, opto-spintronics, and opto-spin-caloritronics [59][60]. Since these vdW materials are flexible and easily stacked together, assembling layers of different vdW magnetic materials, or a vdW material with a non-vdW material, can create novel heterostructures with atomically sharp interfaces and properties that would otherwise be absent in their individual components [41]. Interestingly, large EB effects have recently been observed in FM/AFM vdW interfaces such as oxidized $Fe_3GeTe_2$ [61], $Fe_3GeTe_2/CrCl_3$ [16], $Fe_3GeTe_2/FePS_3$ [62], $Fe_3GeTe_2/MnPX_3$ ($X$=S, Se) [17][18], and $Fe_3GeTe_2/Ir_{22}Mn_{78}$ [63], as well as in defective magnet (DM) vdW interfaces such as $VSe_2/MoS_2$ [64] and FM/vdW DM interfaces such as $Fe/MoS_2$ [65][66]. Owing to the reduced dimensionality, stacking and mechanical flexibility, and tunability of their EB field, these exchange-biased vdW heterostructures appear to be novel platforms for probing low-dimensional vdW nanomagnetism and for developing a new generation of vdW spintronic nanodevices. However, the physical origins of the EB and its association with other effects, such as the MP effect [62][67][68][69], are not fully understood. The following fundamental, important, and unanswered questions are emerging:

i.   *What is the exact coupling mechanism between spins at the FM/AFM interface in FM/AFM vdW systems?*

ii.  *Does the EB effect in FM/AFM vdW systems originate from the direct FM/AFM interfacial coupling?*

iii. *What do FM and AFM spins play roles in the EB effect in FM/AFM vdW systems?*

iv.  *Does type of AFM ordering affect the EB behavior in FM/AFM vdW systems?*

v.   *Is there a thickness limit of the AFM layer for inducing the EB effect in FM/AFM vdW systems?*

vi.  *Can an intermediate magnetically disordered phase be formed at the FM/AFM interface and induce the EB effect in FM/AFM vdW systems?*

vii. *Can FM/SG or AFM/SG interfaces be formed during fabrication and induce the EB effect in vdW magnetic systems?*

*viii.* *What is the role of magnetic proximity of the FM layer on the EB effect in FM/AFM vdW and related systems?*

*ix.* *Does charge transfer play a role in mediating the magnetism and the EB effect in FM/AFM vdW systems?*

*x.* *What is the role of interlayer spin coupling on the EB effect in FM/AFM vdW systems?*

*xi.* *Can the EB effect be tuned to occur at room temperature in vdW magnetic systems?*

*xii.* *Can the EB effect be tuned by external stimuli (electrical field, strain) in vdW magnetic systems?*

This review aims to provide a thorough analysis on EB and related phenomena in recently reported vdW heterostructures and related systems. Emphasis will be placed on an intrinsic vdW ferromagnet $Fe_3GeTe_2$ and its heterostructures (Fig. 1). Effects of oxidization, capping layer, and substrate on the magnetism and EB in $Fe_3GeTe_2$ and its heterostructures are found significant and need careful attention [70][71][72][73][74]. We raise several important questions and propose approaches to addressing those questions. Perspective applications of these and related heterostructures in the emerging fields of spintronics, opto-spintronics, spin-caloritronics, opto-spin-caloritronics, valleytronics, and quantum computation are assessed and proposed. Finally, concluding remarks and an outlook for the future research in this exciting and ever rapidly expanding research field are laid out.

## 2. Exchange Bias and Related Phenomena

### 2.1. Weakly Interlayer-Coupled vdW Ferromagnets

Among the recently discovered 2D intrinsic ferromagnets [75], $Fe_3GeTe_2$ appears to be one of the most promising candidates for spintronics applications [6][52][76][77][78][79][80] as it orders ferromagnetically at high Curie temperature ($T_C$ ~207 K in bulk and ~130 K in monolayer), which can be enhanced to above room temperature upon ionic liquid gating (e.g., $T_C$ ~310 K in trilayer [76]). $Fe_3GeTe_2$ is a layered material, and its lattice and spin structure are illustrated in Fig. 2a. This material belongs to the P63/mmc space group with one $Fe_3Ge$ layer sandwiched by two Te layers. The separation between two adjacent monolayers is ~2.95 Å. The valence states of $Fe_3GeTe_2$ can be viewed as $(Te^{2-})(Fe_I^{3+})[(Fe_{II}^{2+})(Ge^{4-})](Fe_I^{3+})(Te^{2-})$ with inequivalent $Fe_I^{3+}$ and $Fe_{II}^{2+}$ sites within the $Fe_3Ge$ plane [76]. Bulk $Fe_3GeTe_2$ often possesses a large out-of-plane magnetic anisotropy with respect to the vdW planes (Fig. 2a), and the significant strength of this anisotropy stabilizes long-range ferromagnetic order in the monolayer limit. It has been reported that mechanically exfoliated crystals of $Fe_3GeTe_2$ (monolayer and a few layers) have their bulk magnetic characteristic preserved [52][76]. The Curie temperature of $Fe_3GeTe_2$ varies significantly with thickness when the number of layers is less than ~10 but remains almost unchanged for thicker films. However, the $T_C$ of this material in the 2D limit can be largely increased by ionic liquid gating [76]. Unlike $CrI_3$ [45] and $MnSe_2$ [50], $Fe_3GeTe_2$ exhibits relatively weak interlayer exchange coupling [52] and hosts a large Berry curvature responsible for its anomalous Hall effect [81].

It is of practical importance to recall that $Fe_3GeTe_2$ can be easily oxidized when exposed to air, so particular attention must be paid while characterizing the magnetic properties of this material [70]. This matter has also urged researchers to investigate effects of oxidization on the magnetism in mechanically exfoliated crystals of $Fe_3GeTe_2$. Recently, Gweon *et al.* reported a robust and sizable EB effect in $Fe_3GeTe_2$ crystals when the flakes were mechanically exfoliated and annealed in air at 100 °C for 30 min [61]. Annealing was performed to facilitate the formation of an oxide layer on the FM layer (Fig. 2b). The authors hypothesized that this oxide layer would be antiferromagnetic and its interfacial coupling with the pre-existing FM layer would induce the EB effect (Fig. 2c,d). The authors observed a similar temperature dependence of EB field ($H_{EB}$) and coercive field ($H_C$) for annealed $Fe_3GeTe_2$ samples with two different thicknesses (17 and 38 nm, in Fig. 2d), and that the value of $H_{EB}$ remained large even at a FM thickness of ~100 nm. Their observations seem to suggest that the EB effect occurs mainly at the interface between the oxide layer and the outer FM layer while the interior FM layer played a minor role, which is a reasonable assumption given the weak interlayer magnetic exchange interaction between vdW FM layers within the $Fe_3GeTe_2$ film. This intriguing feature distinguishes exchange-biased vdW magnetic heterostructures from conventional non-vdW magnetic bilayers exchange [10].

However, the underlying origin of the observed EB effect in oxidized $Fe_3GeTe_2$ remains elusive. As noted above, an FM/AFM interface does not necessarily induce an EB effect. In annealed $Fe_3GeTe_2$ samples, the magnetic nature of the oxide phase is unknown (i.e., *What are the chemical composition and the thickness of this oxide layer?*), and whether this phase is magnetically ordered or comprises disordered spins analogous to "spin clusters" or SGs [71][72][82][83]. It should also be noted in Fig. 2b that the oxide layer was not uniform over the surface of the FM layer, and the presence of the $AlO_2$ capping layer on top of the oxide layer could also affect the magnetic property of the oxide layer and hence the EB effect. Indeed, the presence of a Pt top layer or $WTe_2$ on $Fe_3GeTe_2$ has been shown to alter the magnetic domain structure of $Fe_3GeTe_2$ [71][72]. Energy Dispersive Spectroscopy (EDS) mapping revealed a strong variation in Fe content across the FM/oxide interface suggesting a possible formation of a less magnetically ordered or magnetically disordered $Fe_{3-x}GeTe_2$ phase in addition to an iron oxide phase (e.g., FeO, $Fe_3O_4$, $\gamma$-$Fe_2O_3$), which could also be formed in $Fe_3GeTe_2$ samples during annealing [70]. The EB effect could thus occur at the $Fe_3GeTe_2$/$Fe_{3-x}GeTe_2$ interface, and the iron oxide layer could act as a pinning layer. These open questions warrant further systematic studies. For example, it is essential to anneal $Fe_3GeTe_2$ at different temperatures and/or for different times to vary the thickness of the oxide layer and investigate its influence on the magnetic coupling between the FM and oxide layers and, consequently, the EB effect. It would be valuable to know at what thickness of oxide layer the EB effect disappears as this could explain the absence of an EB effect in unannealed $Fe_3GeTe_2$ [61] whose oxide layer could be too thin to result in a strong exchange coupling between the FM and oxide layers. In this case, thicker oxide layers may allow a probe of the chemical composition and spin structures in the oxide phase ($Fe_{3-x}GeTe_2$ and/or iron oxide phase) by means of Mössbauer, x-ray magnetic circular dichroism (XMCD), neutron diffraction (ND), scanning tunneling microscopy (STM), and spin-polarized STM measurements to grant deeper insight into the magnetic interfacial coupling mechanism and EB origin. Interfacing a clean (non-oxidized) $Fe_3GeTe_2$ layer with a thin iron oxide layer (~2-5 nm) to create a $Fe_3GeTe_2$/iron oxide heterostructure would also be interesting to understand the role of the iron oxide phase ($FeO$, $Fe_3O_4$, $\gamma$-$Fe_2O_3$), if present, on the magnetism and the EB effect in oxidized $Fe_3GeTe_2$ samples, as well as in its heterostructures, as will be discussed below.

### 2.2. Ferromagnetic vdW/Antiferromagnetic vdW Interfaces

To explore the EB effect in FM/AFM vdW heterostructures, Zhu *et al.* stacked the FM $Fe_3GeTe_2$ with the AFM $CrCl_3$ with different thicknesses of $CrCl_3$ [16]. $CrCl_3$ is an insulating antiferromagnet with a Néel temperature of ~20 K. It is a layered material exhibiting an in-plane magnetic anisotropy (Fig. 3a) [84]. The advantages of this material include easy exfoliation and chemical robustness, which enable a systematic study of the effect of varying thickness of $CrCl_3$ on the magnetism and EB effect in $Fe_3GeTe_2$/$CrCl_3$ heterostructures [16]. Anomalous Hall effect (AHE) measurements (Fig. 3c) revealed sizeable EB effects in these heterostructures (Fig. 3d,e). Relative to bare $Fe_3GeTe_2$ (thickness, 30 nm), the $Fe_3GeTe_2$ (30 nm)/$CrCl_3$ (15 nm or 45 nm) heterostructures were reported to exhibit an enhanced coercivity that increased significantly when the thickness of the CrCl₃ flake increased from 15 to 45 nm (Fig. 3e). However, the EB field was observed to be maximal for a critical thickness of $CrCl_3$ (~20 nm) while above and below which the EB field decreased significantly. Indeed, this thickness-dependent $H_{EB}$ trend is rather different from those reported for non-vdW FM/AFM systems [10]. Further, the authors observed an appearance and enhancement of the EB effect at temperatures below the Néel temperature of the AFM $CrCl_3$. The effect of the cooling field ($H_{CF}$) on the EB field was also investigated, which showed a strong suppression of the EB effect for $H_{CF} > 1$ T at which the AFM state was fully converted into the FM state. Accordingly, the origin of the EB effect in $Fe_3GeTe_2/CrCl_3$ heterostructures was attributed to FM/AFM interfacial coupling. As noted above, however, a sizeable EB effect was also observed in oxidized $Fe_3GeTe_2$ [61] and such an oxide layer could be naturally formed on the surface of $Fe_3GeTe_2$ during $Fe_3GeTe_2/CrCl_3$ formation. If the oxide phase was present between the $Fe_3GeTe_2$ and $CrCl_3$ layers and behaved as "spin clusters", then the AFM $CrCl_3$ could act as a pinning layer and the EB mechanism would be different from what was proposed. Another point is that the EB effect disappeared at ~120 K for oxidized $Fe_3GeTe_2$, while it vanished at a much lower temperature of ~20 K for the $Fe_3GeTe_2/CrCl_3$ heterostructure. This suggests that the oxide layer, if it did exist in that heterostructure, would be very thin and its spin configuration could be impacted by the presence of $CrCl_3$, which would alter interfacial spin coupling and hence the EB effect. To further clarify this, it would be essential to investigate the EB effect in a heterostructure composed of an intentionally oxidized (annealed) $Fe_3GeTe_2$ layer and the $CrCl_3$ layer and compare these results with those reported for the oxidized $Fe_3GeTe_2$ sample and the $Fe_3GeTe_2/CrCl_3$ heterostructure. On the other hand, the large enhancement of $H_C$ in the $Fe_3GeTe_2/CrCl_3$ heterostructure [16] relative to bare (oxidized) $Fe_3GeTe_2$ [61] could suggest a significant dual MP effect induced by the FM layer ($Fe_3GeTe_2$) on the AFM layer ($CrCl_3$) and vice versa. MP and its impact on magnetism and EB should thus be systematically investigated in these heterostructures. In addition, charge transfer between $Fe_3GeTe_2$ and $CrCl_3$ could give rise to the enhanced magnetism in $Fe_3GeTe_2$ and hence the EB effect in the $Fe_3GeTe_2/CrCl_3$ heterostructure, which should also be investigated further.

In another study, Zhang *et al.* reported on the large enhancement of coercive field, Curie temperature, and EB effect in $Fe_3GeTe_2/FePS_3$ and $FePS_3/Fe_3GeTe_2/FePS_3$ heterostructures [62]. Unlike $CrI_3$ and $CrCl_3$, $FePS_3$ is an Ising-type AFM material with both intra- and inter-layer AFM ordering (Fig. 4a) [54]. It belongs to the class of cation-ordered, $CdCl_2$-type, low-dimensional layered magnetic materials and is a Mott insulator. Bulk $FePS_3$ undergoes AFM ordering at $T_N \sim 115$ K, which is lower than the Curie temperature of bulk $Fe_3GeTe_2$ ($T_C \sim 230$ K) (Fig. 4b). The formation of FM/AFM $Fe_3GeTe_2/FePS_3$ heterostructures resulted in a large enhancement of magnetism ($T_C$ and $H_C$) in $Fe_3GeTe_2$, as well as the appearance of an EB effect (Fig. 4c,d). Magneto-optic Kerr effect (MOKE) measurements indicated that $T_C$ increased from 150 K for the mechanically thinned $Fe_3GeTe_2$ crystal (thickness, 8.2 nm) to 180 K for its heterostructure $Fe_3GeTe_2/FePS_3$ where the thickness of the $FePS_3$ layer was about 6 nm. The corresponding increase in $H_C$ by 100 % from 2000 to 4000 Oe was also reported. However, it has remained unknown, from the MOKE experiments, if the saturation magnetization ($M_S$) of $Fe_3GeTe_2$ increased when the $Fe_3GeTe_2/CrCl_3$ interface formed. Unlike the case of the $Fe_3GeTe_2/CrCl_3$ interface, a double-shifted hysteresis loop feature was observed for the $Fe_3GeTe_2/FePS_3$ heterostructure (Fig. 4d). Interestingly, relative to the $Fe_3GeTe_2/FePS_3$ heterostructure (one interfacial magnetic coupling), the magnetism ($T_C$ and $H_C$) and EB effect were further enhanced in the $FePS_3/Fe_3GeTe_2/FePS_3$ heterostructure (double interfacial magnetic coupling) in which an additional AFM layer of $FePS_3$ (the same thickness) was stacked on the other surface of the $Fe_3GeTe_2$ flake (Fig. 4e,f). In this tri-layer heterostructure, a step-like magnetization behavior in the *M-H* loop was also observed (Fig. 4f). The origins of the enhanced magnetism and EB effect were attributed to FM/AFM interfacial coupling that was mediated by the MP effect. It was suggested that the presence of the AFM $FePS_3$ layer pinned surface spins within the FM $Fe_3GeTe_2$ layer causing the $H_C$ of the FM layer to increase. However, the MOKE data (Fig. 4c) revealed a noticeable increase in $M_S$ and a decrease in magnetic anisotropy in the $Fe_3GeTe_2/FePS_3$ heterostructure when compared to bare $Fe_3GeTe_2$. While the MP effect indeed plays an important role in mediating the magnetism and EB effect in $Fe_3GeTe_2/FePS_3$ and $FePS_3/Fe_3GeTe_2/FePS_3$ heterostructures, the underlying origins of the observed EB and MP effects remain unclear and warrant further studies. As noted earlier, oxidization of mechanically exfoliated $Fe_3GeTe_2$ flakes could occur during sample preparation and its effect on the magnetism and EB was not examined in that work [62]. To elucidate this, it would be necessary to perform a systematic comparative study of the magnetic and EB properties of both clean (non-oxidized) and oxidized (via annealing) $Fe_3GeTe_2/FePS_3$ heterostructures. Varying $FePS_3$-thickness in such $Fe_3GeTe_2/FePS_3$ heterostructures would also be essential to address the open questions regarding the effects of type of AFM ordering, interlayer coupling, and reduced dimensionality on the magnetism and EB behavior (the double-shifted hysteresis behavior). Another effect that was not considered in the work [62] is "charge transfer" that could occur at the $Fe_3GeTe_2/FePS_3$ interface and give rise to the enhanced magnetism and EB field in this heterostructure.

In an extended study, Dai *et al.* also observed enhanced magnetism and EB effect in FM/AFM $Fe_3GeTe_2/MnPX_3$ ($X$ = Se and S) heterostructures (Fig. 5) [17]. Bulk $MnPS_3$ has a monoclinic structure with the point group of $C_2h$ (2/m). In $MnPS_3$, each Mn atom is surrounded by six S atoms and connected to two P atoms above and below the Mn plane (Fig. 5a). Unlike bulk $FePS_3$, bulk $MnPS_3$ exhibits Heisenberg-type AFM ordering with a Néel temperature of 78 K [85]. The magnetic moments of Mn in $MnPS_3$ are ferromagnetically coupled across the vdW gap to its inter- and intra-layer nearest neighbor. Although bulk MnPS$_3$ and MnPSe$_3$ share a similar crystal structure (Fig. 5a), their magnetic characteristics are quite different, especially at low temperatures (Fig. 5b,c). Bulk MnPSe$_3$ exhibits Ising-type AFM ordering with a $T_N$ of 74 K. The magnetic moments of Mn in this material are antiferromagnetically coupled to its inter- and intra-layer nearest neighbor, which is different from what was observed for bulk MnPS$_3$. While MnPS$_3$ possesses dominant out-of-plane magnetic moments, neutron scattering measurements have revealed that the majority of spins in MnPSe$_3$ are in-plane. Spin-orbit coupling has been found to be negligible in the former but significant in the latter. These differences could allow us to interpret different magnetic behaviors and the resulting EB effects between these two systems when each are stacked with Fe$_3$GeTe$_2$ to form corresponding Fe$_3$GeTe$_2$/MnPS$_3$ and Fe$_3$GeTe$_2$/MnPSe$_3$ heterostructures (Fig. 5b-d). Given that the thickness of each flake (Fe$_3$GeTe$_2$, MnPS$_3$, and MnPSe$_3$) was 23 nm to form the heterostructures, the flakes on their own should exhibit bulk magnetic characteristics i.e., MnPS$_3$ and MnPSe$_3$ should order antiferromagnetically below 80 K. This appears to be in line with the strong development of the EB effect observed just below this temperature in the Fe$_3$GeTe$_2$/MnPS$_3$ and Fe$_3$GeTe$_2$/MnPSe$_3$ heterostructures (Fig. 5d). As compared to Fe$_3$GeTe$_2$/MnPS$_3$, larger values of $H_C$ and $H_{EB}$ were reported for Fe$_3$GeTe$_2$/MnPSe$_3$ at temperatures below ~120 K. This could be attributed to spin-orbit coupling that is more significant in MnPSe$_3$ than MnPS$_3$. Nonetheless, the two-step magnetization versus temperature behavior observed around 100 K in the case of MnPSe$_3$ remains questionable as a re-orientation of spins at the Fe$_3$GeTe$_2$/MnPSe$_3$ interface may occur around this temperature. This hypothesis must be examined and validated experimentally. There are also open questions about the effects of oxidization (the oxidized Fe$_3$GeTe$_2$ layer), interlayer coupling within the AFM layer, charge-transfer across the Fe$_3$GeTe$_2$/MnP$X_3$ ($X$ = S, Se) interface, and the effects of reduced dimensionality on the magnetism of $Fe_3GeTe_2$ and hence the EB effect observed in these heterostructures [18].

In a recent study, Fu *et al.* predicted EB and quantum anomalous Hall (QAH) effects in $CrI_3$/$MnBi_2Te_4$ FM/AFM heterostructures [86]. Recall that bulk $MnBi_2Te_4$ is an A-type topological AFM that exhibits a quantized Hall resistance when an applied external magnetic field exceeds the critical field of the spin-flop transition (> 6 T) [86]. Thus, strong interfacial coupling between the FM monolayer of $CrI_3$ and the AFM $MnBi_2Te_4$ layer was predicted to induce the EB effect, which could change electronic states near the Fermi level. The EB mechanism would then originate from the long $Cr$-$e_g$ orbital tails that are strongly hybridized with Te $p$ orbitals and an out-of-plane surface magnetic ordering in $MnBi_2Te_4$ would be induced via MP to the FM $CrI_3$ monolayer. These interesting predictions should be validated experimentally. In this regard, it would also be interesting to interface $MnBi_2Te_4$ or similar [87] [88] with other types of vdW magnets such as $Fe_3GeTe_2$ or $Fe_5GeTe_2$, and further explore EB and QAH effects in those heterostructures.

### 2.3. Ferromagnetic vdW/Antiferromagnetic non-vdW Interfaces

Exchange biased non-vdW magnetic systems have been extensively explored for applications in spintronic devices based on spin-orbit torque (SOT)-driven magnetization switching [89]. Owing to their atomically flat surfaces and 2D magnetic properties, there has been a growing interest in exploring vdW magnetic materials for use in SOT-based spintronic devices [5][6]. Alghamdi *et al.* created heterostructures consisting of 5-nm Pt films sputtered onto the surfaces of ~20 nm exfoliated $Fe_3GeTe_2$ flakes and measured the second harmonic Hall responses as the applied magnetic field rotated the magnetization of $Fe_3GeTe_2$ in the plane [90]. They achieved a large SOT efficiency and attributed it to the atomically flat $Fe_3GeTe_2$/Pt interface. The SOT efficiency of the $Fe_3GeTe_2$/Pt heterostructure is comparable with that of the best heterostructures comprising non-vdW ferromagnetic metals and is greater compared to those based on non-vdW ferrimagnetic insulators, which makes this heterostructure an attractive candidate for use in highly efficient spintronic nanodevices. By adding an AFM layer of $Ir_{22}Mn_{78}$ (IrMn) in between the $Fe_3GeTe_2$ and Pt layers (Fig. 6a), Zhang *et al.* recently observed both EB and SOT effects in the $Fe_3GeTe_2$/IrMn/Pt heterostructure (Fig. 6b,c) [63]. A large EB effect (up to 895 Oe) was achieved in this heterostructure (Fig. 6b), which is greater than what has been observed in previously reported bilayers including $Fe_3GeTe_2$/$CrCl_3$ [16], $Fe_3GeTe_2$/$FePS_3$ [62], and $Fe_3GeTe_2$/$MnPX_3$ ($X$ = S, Se) [17]. Relative to the $Fe_3GeTe_2$/Pt heterostructure, a larger SOT-driven magnetization switching performance was achieved for the $Fe_3GeTe_2$/IrMn/Pt heterostructure (Fig. 6c). Nonetheless, effects of the oxidization layer of $Fe_3GeTe_2$ on the SOT efficiency in the $Fe_3GeTe_2$/Pt and $Fe_3GeTe_2$/IrMn/Pt heterostructures, as well as on the large EB coupling in the $Fe_3GeTe_2$/IrMn/Pt heterostructure, have remained questionable. It would thus be essential to investigate this effect in both clean and oxidized $Fe_3GeTe_2$/Pt heterostructures by annealing the $Fe_3GeTe_2$ flakes to form an oxidization layer on their surfaces before depositing the top Pt layer. Other vdW magnetic systems have recently been explored for SOT-based device applications [5][6] and further studies are needed to examine if interfacing these vdW magnets with other materials can enhance the SOT efficiency and the EB effect, as well as utilize their combined functionalities for modern spintronics.

### *2.4. Ferromagnetic vdW/Ferromagnetic non-vdW Interfaces*

While most recently discovered intrinsic 2D vdW magnets exhibit ferromagnetic ordering well below room temperature and possess a rather weak magnetic signal in the 2D limit, interfacing a 2D vdW magnet with a 3D non-vdW ferromagnet has proved an alternative and effective approach to enhance the magnetic property of a 2D vdW magnet via MP induced by the 3D non-vdW ferromagnet and vice versa [67]. However, a clear understanding of the magnetic interfacial coupling mechanism in these hybrid structures is lacking. Chen *et al.* deposited a 6-nm Ni layer on exfoliated monolayer and bilayer $Fe_3GeTe_2$ flakes (and also on an exfoliated $Cr_2Ge_2Te_6$ flake) to create $Fe_3GeTe_2/Ni$ (and $Cr_2Ge_2Te_6/Ni$) heterostructures and investigated their magnetic properties using spin torque ferromagnetic resonance (FMR) [67]. It is worth mentioning that relative to the bare Ni film, the presence of the $Fe_3GeTe_2$ or $Cr_2Ge_2Te_6$ layer increased the perpendicular magnetic anisotropy (PMA) and magnetic damping of the adjacent Ni film. The PMA effect became stronger when the $Fe_3GeTe_2$ thickness increased from monolayer to bilayer highlighting the important role of interlayer magnetic coupling. The observation of such an enhanced effect at room temperature also suggests that the presence of the Ni layer could significantly enhance the magnetic ordering within $Fe_3GeTe_2$ or $Cr_2Ge_2Te_6$ via their magnetic interfacial coupling despite the fact that the Curie temperatures of $Fe_3GeTe_2$ and $Cr_2Ge_2Te_6$ are well below room temperature. These findings are of practical importance and may provide an effective approach for utilizing 2D magnets in modern spintronic devices that operate at ambient temperature. However, the exact nature of the interfacial spins (i.e., their spin textures) and their couplings to Fe spins located within each layer are unknown in these hybrid systems. The case could become even more complicated if the oxidized $Fe_3GeTe_2$ layer–possibly formed during the exfoliation of $Fe_3GeTe_2$ crystals or during the deposition of the Ni layer on the $Fe_3GeTe_2$ flake–was present in between the $Fe_3GeTe_2$ and Ni layers. Further studies are needed to clarify this effect. Since oxidized $Fe_3GeTe_2$ flakes could exhibit an EB effect, it would be necessary to create oxidized $Fe_3GeTe_2/Ni$ heterostructures and investigate impacts of EB on the magnetic property of the Ni layer, as well as on the spin transport across the $Fe_3GeTe_2/Ni$ interface. Reducing the thickness of the Ni layer to 2D would also allow one to create novel 2D $Fe_3GeTe_2/Ni$ heterostructures with potentially emerging new magnetic functionalities. It has been shown that $MnBi_8Te_{13}$ is an intrinsic topological ferromagnetic insulator whose interlayer interaction and surface state can be modulated upon photoexcitation [91]. Interfacing a 2D vdW magnet like $Fe_3GeTe_2$ or $Fe_5GeTe_2$ with this type of magnetic topological insulator could also create a novel heterostructure whose interfacial magnetism can be tuned by external stimuli such as light. Such heterostructures would be an interesting subject for future research, and the knowledge of which will lay a foundation for the development of modern 2D spintronic devices.

### 2.5. Ferromagnetic vdW/Defective Magnetic vdW Interfaces

2D vdW transition metal dichalcogenides (TMDs), such as $TX_2$ monolayers ($T$ = Mo, W; $X$ = S, Se, Te), are at the heart of many important device applications such as field-effect transistors, photodetectors, photon emitters, spintronics, valleytronics, and quantum computers [3][60][92][93]. Recent studies have shown that the magnetic or magneto-optical property of a non-magnetic TMD can be induced or enhanced by stacking it with a vdW magnet [72][94][95][96]. This is a direct consequence of the MP effect induced by the vdW magnet [37]. Given the fact that the $MoS_2$ monolayer exhibits a large spin splitting and an out-of-plane spin polarization, sandwiching the $MoS_2$ monolayer in between two ferromagnetic (Fe, FeNi, or Co) electrodes has been predicted to promote the spin injection and enhance the magnetoresistance (MR) up to 300% [97]. However, only a 0.4% MR ratio was experimentally achieved in the NiFe/$MoS_2$/NiFe spin valve (SV) structure [98]. This low value could arise from the presence of a non-uniform interface between the non-vdW NiFe and vdW $MoS_2$ layers. To overcome this, Lin *et al.* created a novel vdW interface comprising $Fe_3GeTe_2$ and $MoS_2$ layers and both of which were mechanically exfoliated from their bulk crystal counterparts [99]. They showed that monolayer $MoS_2$ acted as a conducting interlayer in the $Fe_3GeTe_2$/$MoS_2$/$Fe_3GeTe_2$ structure [99], and an enhanced MR ratio of 3.1% was achieved at 10 K in this SV structure (Fig. 7), which is about 8 times greater than that of the NiFe/$MoS_2$/NiFe structure [98]. A large tunneling MR effect (up to 41%) has recently been achieved in a Fe$_3$GeTe$_2$/InSe/Fe$_3$GeTe$_2$ structure. This effect can be attributed to the presence of a pinhole-free tunnel barrier at the Fe$_3$GeTe$_2$/InSe interface. These observations pinpoint the importance of a flat interface, which can only be formed between vdW materials, and the superior advantages of the vdW heterostructures for SV-based spintronics. Nonetheless, there are several open questions that need to be addressed. First, the underlying origin of the MR enhancement in these SV structures is not fully understood. It should be recalled that Fe$_3$GeTe$_2$ flakes could become oxidized on their surface during mechanical exfoliation, so it is essential to examine the effect of Fe$_3$GeTe$_2$ oxidization on the spin injection and hence MR behavior in Fe$_3$GeTe$_2$/MoS$_2$/Fe$_3$GeTe$_2$ [99] and Fe$_3$GeTe$_2$/InSe/Fe$_3$GeTe$_2$ structures [100]. It remains unclear if the asymmetry observed in the MR curves was caused by the presence of the oxidized surface layer of Fe$_3$GeTe$_2$ that could be antiferromagnetically coupled to the interior layer of Fe$_3$GeTe$_2$. It is also essential to examine if the MR enhancement in these SV structures could be driven by the MP effect of the ferromagnetic Fe$_3$GeTe$_2$ layer onto the MoS$_2$ monolayer. While a defect-free MoS$_2$ monolayer is known to be diamagnetic in nature, in practice, most CVD-grown MoS$_2$ monolayers have been reported to exhibit defect-induced weak ferromagnetic ordering even at room temperature due to the presence of S and Mo vacancies [101]. This means that magnetic interfacial coupling between Fe$_3$GeTe$_2$ and MoS$_2$ could be significant, thus affecting the MR behavior in this SV structure. In the context of the MP effect, it is very important to know how the magnetic domain structure of Fe$_3$GeTe$_2$ varies when Fe$_3$GeTe$_2$ is placed in close contact with a 2D-TMD like monolayer MoS$_2$. Indeed, Lorentz transmission electron microscopy measurements on the Fe$_3$GeTe$_2$/WTe$_2$ heterostructure have evidenced the effect of a WTe$_2$ bilayer on the magnetic domain structure of the mechanically exfoliated Fe$_3$GeTe$_2$ flake [72]. The labyrinth-like magnetic domain structure observed in the Fe$_3$GeTe$_2$ flake (30 layers) was transformed to an aligned and stripe-like magnetic domain structure when the Fe$_3$GeTe$_2$ flake was interfaced with the WTe$_2$ layer. Such a difference in the magnetic domains identifies a significant interfacial coupling between the $Fe_3GeTe_2$ and $WTe_2$ layer that contributes to the strong Dzyaloshinskii-Moriya interaction (DMI) in the $Fe_3GeTe_2/WTe_2$ heterostructure [71]. In this case, spin orbit coupling proximity could play an important role. Since $Fe_3GeTe_2$ is a metal, and $MoS_2$ a semiconductor, the conductivity mismatch between these two materials could promote charge transfer across the $Fe_3GeTe_2/MoS_2$ interface thus mediating the overall magnetism of the $Fe_3GeTe_2/MoS_2/Fe_3GeTe_2$ heterostructure. Meanwhile, charge transfer-mediated magnetism has recently been reported in several vdW and non-vdW magnetic heterostructures [102]. Zhong *et al.* demonstrated that the spin-dependent charge transfer between monolayer $WSe_2$ and bi/trilayer $CrI_3$ in $WSe_2/CrI_3$ heterostructures is dominated by the interfacial $CrI_3$ layer due to which the MP effect impacts the magnetic and magneto-optic response of the heterostructure differently [41]. These hypotheses need experimental and theoretical validation.

### *2.6. Ferromagnetic non-vdW/Defective Magnetic vdW Interfaces*

Although it has remained challenging to form uniform interfaces between a non-vdW magnetic material and a non-magnetic or defective magnetic vdW materials like TMDs, a large body of work on these heterostructures has been reported [66][103][104][105]. This challenge could arise from the fact that the surface of a non-vdW magnetic material is quite rough (surface roughness of a few nm) compared the thickness of a vdW monolayer (less than 1 nm). DFT calculations performed on a 2D $Fe/MoS_2$ heterostructure have shown that the Fe atoms on $MoS_2$ are ferromagnetically coupled with each other with a magnetic moment of ~2.0 $\mu_B$/atom [66]. Monte Carlo simulations with Heisenberg spin Hamiltonians have predicted a high Curie temperature of 465 K, and this 2D $Fe/MoS_2$ heterostructure exhibits a half-metallic characteristic. However, these predictions have not been experimentally validated to date. It is also important to mention that the conductivity mismatch between the metal (Fe) and the semiconductor ($MoS_2$) could promote charge transfer across the Fe/MoS$_2$ interface and the Fe layer could magnetize the MoS$_2$ monolayer via the MP effect–both of which were not included in these calculations. Effect of defects (e.g., S and Mo vacancies) on the magnetism of MoS$_2$ and the magnetic interfacial coupling between the Fe and MoS$_2$ layers were also not examined. The Fe/MoS$_2$ heterostructures with different thicknesses of Fe layers were hydrothermally synthesized and a noticeable EB effect was observed to occur at low temperatures in these samples [65]. The authors suggested a possible charge transfer from Fe to MoS$_2$ that induced magnetic moments that are antiferromagnetically coupled to the Fe spins. However, the mechanism of charge transfer that mediates the magnetism and the origin of the EB effect in these systems are not fully understood. Due to the nature of Fe/MoS$_2$ synthesis, Fe atoms could also be doped into the MoS$_2$ lattice, which would mediate the magnetism of MoS$_2$ and therefore give rise to the observed EB effect. This could be possible indeed as Fe-doped MoS$_2$ monolayers have been reported to show ferromagnetic ordering above room temperature [58]. In a relevant study, Wang *et al.* showed a strong hybridization between Mo/S atoms with Ni/Fe atoms in the FeNi/MoS$_2$ heterostructure and its significant impact on the electrical transport and spin transport in this bilayer [98]. The SV effect was observed at temperatures as high as 240 K with the largest MR ratio of 0.73% achieved at 20 K. This value of MR is about 9 times smaller than what was theoretically predicted (9%) for the same system. Such a large discrepancy between experiment and theory suggests other possible contributions (surface roughness, magnetic proximity, atomic diffusion, doping) to the magnetism, charge transport, and spin transport in the FeNi/MoS$_2$ system. Magnetic element specific x-ray photoemission electron microscopy (PEEM) of a Co/MoS$_2$ heterojunction revealed the formation of micron-sized magnetic domains on monolayer MoS$_2$ [106]. Meanwhile, x-ray photoelectron spectra evidenced the charge transfer from Co to S at the Co/MoS$_2$ interface. Relative to the Fe/MoS$_2$ heterostructure, a different ferromagnetic behavior was observed for the Co/MoS$_2$ heterostructure, which could be attributed to a different orbital hybridization effect at the Co/MoS$_2$ interface. This could also explain the enhanced SV effect (the largest MR ratio of 8% at 4 K) reported for the Co/WS$_2$ heterostructure [103] as well as the enhanced spin Seebeck effect (SSE) in the Ni$_{81}$Fe$_{19}$/WS$_2$/Pt heterostructure [107]. On the other hand, a giant MR effect (~30%) has been observed in the MoS$_2$ monolayer upon its direct contact with the ferrimagnetic substrate YIG in the YIG/MoS$_2$ heterostructure [104]. This value of MR is one order of magnitude greater than that reported for the MoS$_2$/CoFe$_2$O$_4$ heterostructure. The giant MR effect has been attributed to interfacial spin accumulation due to the YIG layer. Using spin-resolved photoluminescence spectroscopy (SRPS) and magnetic circular dichroism (MCD), Tsai *et al.* revealed room temperature ferromagnetic ordering in MoS$_2$ induced by the MP effect of YIG [94], and that the MP-induced spins are antiferromagnetically coupled to the YIG spins (Fig. 8). Charge transfer has also been proposed to occur at the YIG/MoS$_2$ interface and act as the mediator of the magnetism in the heterostructure. A large enhancement of the SSE has also been reported in YIG/monolayer WSe$_2$/Pt heterostructures [108]. The MP effect of YIG on the WSe$_2$ monolayer magnetism has been shown to impact spin transport and hence the spin-to-charge conversion efficiency in YIG/monolayer WSe$_2$/Pt [105]. These studies yield further insights into spin manipulation in non-vdW magnet/vdW 2D-TMD heterostructures, but also raise open questions regarding the underlying origins of MP-mediated interfacial magnetism.

### 2.7. Defective Magnetic vdW Interfaces

Although there is an ongoing debate on the origin of the observed ferromagnetism in metallic VSe$_2$ monolayers [49][109][110][111][112][113][114][115], recent studies have suggested both intrinsic and extrinsic sources of magnetism in these films [109][113][115]. While decoupling intrinsic magnetism from defect-induced (extrinsic) magnetism is a challenging task, the latter contributes more significantly to the magnetism of this 2D system [109][115]. Because bulk $MoS_2$ exhibits, in addition to its diamagnetic background, a relatively weak ferromagnetic ordering at RT due to S and Mo vacancies or defects, its magnetism and interfacial coupling with the $VSe_2$ layer have been shown to enhance the saturation magnetization and coercivity of the $VSe_2/MoS_2$ system (Fig. 9a) as compared to the $VSe_2/HOPG$ system in which HOPG is purely diamagnetic [49]. The large enhancement of $M_S$ and $H_C$ in $MoS_2/VSe_2$ (Fig. 9b) due to the exchange interaction between the $VSe_2$ and $MoS_2$ layers raises an important question: *Can this FM/FM interaction lead to an exchange bias effect?*

Kalappattil *et al.* recently observed a sizable EB effect in the monolayer $VSe_2$/single crystal $MoS_2$ heterostructure (see Fig. 9b,c), which is absent in the case of the $VSe_2/HOPG$ system [64]. The EB field as large as 310 Oe was achieved at 10 K. It is worth mentioning that the enhancements of both saturation magnetization ($M_S$) and coercivity ($H_C$) in the field-cooled *M-H* loop compared to the zero-field-cooled *M-H* loop at low temperatures (see, for example, Fig. 9b). This indicates that the cooling magnetic field could strengthen the MP-driven interfacial coupling between the $MoS_2$ and $VSe_2$ layers. The magnetic field enhanced MP effect could also facilitate long-range FM ordering within each layer. Since the $MoS_2$ film was not fully covered by the $VSe_2$ layer, randomly oriented $MoS_2$ spins (spin clusters) that were not pinned by the magnetic region in $VSe_2$ could be rotated towards the external magnetic field direction. As a result, the saturation magnetization and coercivity values were enhanced in the $VSe_2/MoS_2$ heterostructure relative to its individual components [49]. Such a feature is interesting and rather different from that observed in FM/AFM non-vdW bilayers for which AFM susceptibility is zero resulting in no coercivity enhancement [10]. Fig. 9c shows the temperature dependence of $H_{EB}$ and $H_C$. As temperature decreases, $H_C$ started increasing as expected for a typical FM system. However, at temperatures below 100 K, $H_C(T)$ exhibited a slope change. An exponential increase in $H_{EB}$ below 50 K is noticeable, and this could be attributed to the establishment of long-range ferromagnetic ordering in MoS₂. Recently, Guguchia *et al.* demonstrated from muon spin rotation spectroscopy (mSR) and scanning tunneling microscopy (STM) measurements that $MoX_2$ ($X$ = Se, Te) possesses low-temperature FM ordering (~50-100 K) arising from intrinsic Mo vacancies [116]. This, along with the MP effect of VSe₂ on the MoS₂ layer, led to a very strong exchange energy, which, in turn, caused the EB field and coercivity to increase in the VSe₂/MoS₂ heterostructure at low temperatures. The cooling field ($H_{FC}$) dependence of the EB field ($H_{EB}$) taken at 10 K indicated that $H_{EB}$ first increased with an increase in cooling field, reached a maximum at a certain value of $H_{FC}$ (~1 T), and finally decreased for larger values of $H_{FC}$. Such behavior is typical in SG/FM systems and shows the complex magnetic coupling and spin frustration at the VSe₂/MoS₂ interface. The observation of the sizeable EB effect in defective magnetic vdW TMD interfaces such as VSe₂/MoS₂ is surprisingly interesting, which once again points to the fact that the FM/AFM interface is not the only condition for inducing the EB effect. The presence of a FM/SG interface or the coexistence of the FM and spin frustration could also lead to the EB effect. Nonetheless, several questions remain to be answered. For instance, it is unclear how spins of V atoms and spins induced by defects (Se vacancies) are coupled to MoS₂ spins (either Mo- or S-vacancy induced) at VSe₂/MoS₂ interfaces and contribute to the EB effect. Recall that in the VSe₂/MoS₂ system, the presence of the VSe₂ layer with a larger work function of ~4.5 eV (as compared to ~4.1 eV for the MoS₂ layer) leads to an accumulation of electrons in the VSe₂ layer and creates a depleted region in the MoS₂ side of the hetero-interface and the subsequent formation of a Schottky barrier. The accumulation of electrons in the VSe₂ layer could give rise to the enhanced ferromagnetism in this layer and, hence, contributes to the change in net magnetization of the VSe₂/MoS₂ heterostructure and hence the observed EB effect. Further studies are needed to verify this hypothesis. It would also be interesting to investigate how the interfacial magnetism and EB behavior in the $VSe_2$/$MoS_2$ heterostructure are altered when the thickness of the $MoS_2$ film is reduced to a few layers or the monolayer limit. Since the net magnetization of the $VSe_2$ film can be tuned by controlling Se-vacancy concentration [109], it would also be interesting to examine if the EB field of the $VSe_2$/$MoS_2$ heterostructure could be tuned by varying the magnetization of the $VSe_2$ layer. Given the fact that other TMDs, such as Se-deficient $WSe_2$ and $TiSe_2$ monolayers, could exhibit ferromagnetic ordering even at room temperature, it would be a worthwhile endeavor to explore the interfacial magnetism and EB effect in $WSe_2$/$MoS_2$ and $TiSe_2$/$MoS_2$ heterostructures. Engineering of atomic defects for tunable interfacial magnetism and EB would also offer a promising platform for 2D van der Waals spintronics [117].

## 3. Perspective applications

Magnetic anisotropy plays a crucial role in the applications of magnetic materials in spintronic devices. In magnetic heterostructures, tailoring exchange magnetic anisotropy (also known as interface anisotropy) has proved its usefulness in improving magnetic functionality and hence the performance of a spintronic device [18][63] [77][83] [118][119] [120][121][122]. We highlight below the important effects that MP-mediated exchange magnetic anisotropy may have on spin configurations, spin-transport, thermo-spin-transport, and valleytronic properties of 2D van der Waals magnet-based heterostructures, which are open to a wide range of important technological applications (Fig. 11).

### 3.1 Spintronics and opto-spintronics

Spin valve structures form the basis of spintronic devices. In these magnetic junctions, the quality of interfaces between the ferromagnetic electrodes and the spacing layer plays a decisive role in the magnetoresistance effect [3]. While poor interfaces are main concerns for non-vdW material-based SV devices, this issue can be overcome by using vdW materials because of their atomically flat interfaces. Indeed, Song *et al.* have shown a giant tunneling magnetoresistance effect (~19,000%) at low temperatures in a novel vdW heterostructure in which an atomically thin $CrI_3$ was sandwiched between graphene contacts [123]. Zhou *et al.* theoretically demonstrated an extremely large tunneling magnetoresistance of 846% at room temperature in a vdW $MoS_2$/$VSe_2$/$MoS_2$ magnetic tunneling junction (MTJ) [124]. Because of the strong spin Hall conductivity of $MoS_2$, SOT is appropriate for magnetization switching of the $VSe_2$ layer. However, this prediction needs experimental validation. Large SV effects have been experimentally achieved in $Fe_3GeTe_2$-based vdW systems such as $Fe_3GeTe_2$/$MoS_2$/$Fe_3GeTe_2$ [99], $Fe_3GeTe_2$/$WS_2$/$Fe_3GeTe_2$ [125], $Fe_3GeTe_2$/$InSe$/$Fe_3GeTe_2$ [126], $Fe_3GeTe_2$/h-BN/$Fe_3GeTe_2$[127], and $Fe_3GeTe_2$/graphite/$Fe_3GeTe_2$ [128]. These proof-of-concept devices demonstrate a promising perspective towards proximitized vdW materials for use in energy-efficient 2D vdW spintronics.

Owing to atomically flat vdW interfaces, the highly efficient SOT and magnetization switching of $Fe_3GeTe_2$, $Cr_2Ge_2Te_6$, and $CrI_3$ have also been achieved [79][63][90][129][130][131]. Alghamdi *et al.* have shown that the SOT efficiency achieved in $Fe_3GeTe_2$/Pt is comparable with that of the best heterostructures containing 3D ferromagnetic metals and much larger than that of heterostructures containing 3D ferrimagnetic insulators [90]. It is anticipated that proximitized vdW heterostructures may also provide better electrical manipulation and notable thermal perturbations that allow a lower analytical current for switching magnetization as compared to non-vdW magnetic junctions in SOT-mediated magneto-resistive memory devices [79][132][133][134]. While the surface magnetism of ferromagnetic electrodes is important in both SV and SOT systems, recall the significant effect of oxidization on the surface magnetic property of mechanically exfoliated $Fe_3GeTe_2$. In particular, oxidized $Fe_3GeTe_2$ has been shown to exhibit the EB effect [61], and the effect that exchange anisotropy may have on the MR and SOT should be investigated systematically.

Research has shown that it is possible to use light via a femtosecond laser pulse to manipulate the ferromagnetism (both magnetization and magnetic anisotropy values) in 2D magnets and heterostructures [135]. Interestingly, light-driven ferromagnetism has been observed at room temperature in atomically thinned $Fe_3GeTe_2$ [135]. The effect has been attributed to the change in the electronic structure of $Fe_3GeTe_2$ due to optical doping. This important observation reveals a new perspective for optically controlling SV and SOT effects in $Fe_3GeTe_2$-based heterostructures or similar vdW heterostructures at room temperature, which has the potential to establish a new subfield of opto-spintronics. Future studies are needed to realize all of this.

### 3.2 Spin-caloritronics and opto-spin-caloritronics

In a heavy metal (HM)/ferromagnet (FM) bilayer such as $Pt/Y_3Fe_5O_{12}$, the application of an external magnetic field and a temperature gradient in the FM layer can generate a pure spin current, which can then be converted into an electrical voltage via the inverse spin Hall effect (ISHE) due to the strong spin-orbit characteristic of the HM. Such phenomenon is now widely known as the spin Seebeck effect (SSE), which was first discovered in Pt/FeNi bilayers by Uchida *et al*. in 2008 forming the basis for the development of spin-caloritronic devices [136][137]. Recent research has shown the large anomalous Nernst effect (ANE) in $Fe_3GeTe_2$ to be a result of charge current driven by the temperature gradient [138]. It is likely that both SSE and ANE coexist in $Fe_3GeTe_2$, but no study has been made to decouple their relative contributions. In SSE and ANE systems, surface/interface magnetic anisotropy has been shown to play an important role [139][140]. As mentioned earlier, it is possible to manipulate the magnetic anisotropy of $Fe_3GeTe_2$ by light. This suggests a new possibility of optically controlling SSE and ANE in $Fe_3GeTe_2$ and its heterostructures. Further studies should be performed to validate this hypothesis.

On the other hand, the giant enhancement of SSE has been reported in a Pt/YIG bilayer by inserting a semiconducting TMD monolayer of WSe$_2$ [108]. In this context, the addition of a 2D-TMD magnetic semiconductor (e.g., V-doped WS$_2$ or V-doped WSe$_2$ monolayer [57][55][56]) has been proposed to reduce the conductivity mismatch between the HM and FM layers and enhance the spin mixing conductance across the HM/FM interface both resulting in further enhancement of the SSE in Pt/YIG [59]. Since the magnetization of a magnetic TMD monolayer (e.g., V-WS$_2$) has recently been demonstrated to be manipulated by varying the intensity of an irradiated light at low power [141], it is possible to use light to control the SSE in HM/magnetic 2D-TMD/FM heterostructures. This idea sparked a new research subfield named "opto-spin-caloritronics" [59]. In a similar perspective, exploring ultrafast magnetism in 2D-TMD magnets and heterostructures would also provide a novel platform for ultrafast opto-spin-caloritronics.

### 3.3 Valleytronics and spin-valleytronics

Valleytronics combines valley and electronics in a way that exploits local extrema, or "valleys," in the electronic band structure of a semiconductor like a TMD monolayer [3][142][143]. This effect has been observed over a wide range of 2D-TMD semiconductors such as monolayers of WSe$_2$, WS$_2$, and MoS$_2$ [144][145][146]. The MP effect has recently been reported to enhance valley Zeeman splitting in these 2D-TMDs by interfacing them with magnetic materials that can fulfill the increasing requirements of spin-valleytronics [104][143][147][148]. Theoretical studies show that the magnetization can be induced in monolayer MoS$_2$ when interfaced with a magnetic metal like Ni [149] or a ferromagnetic TMD like EuS [150]. Recently, Zhao *et al.* experimentally demonstrated the use of a magnetic exchange field (MEF) from an EuS substrate to enhance valley splitting in monolayer WSe$_2$ [151]. More recently, Seyler *et al.* optically manipulated the magnetization of an ultrathin ferromagnetic insulator CrI$_3$ to tune MEF over a range of 20 T, thus enabling a continuous control of the magnitude and sign of valley polarization and Zeeman splitting in monolayer $WSe_2$ [152]. However, these vdW magnets (EuS, $CrI_3$) order ferromagnetically at low temperature (below 100 K) hindering valleytronic devices from operating at room temperature. To realize their practical applications, the valley spin states should be controllable by magnetic field at room temperature. In this context, Kim *et al.* recently demonstrated that the interfacial anisotropy in a $Fe_3GeTe_2$/$WSe_2$ heterostructure can be controlled by varying adjacent layers of $Fe_3GeTe_2$, which leads to an appearance of multiple magnetic behaviors in a single channel [83]. This system is of potential interest for exploring spin valleytronics.

As a promising alternative, magnetic doping into 2D-TMDs has proved its usefulness in enhancing the valley Zeeman splitting at room temperature [153]. Li *et al.* demonstrated an enhanced valley Zeeman splitting at 300 K ($g_{eff}$ = -6.4) in a Fe-doped $MoS_2$ monolayer relative to pristine $MoS_2$ [153]. The $g_{eff}$ factor can also be tuned to -20.7 by increasing the Fe concentration, which has been attributed to the enhanced Heisenberg exchange interaction of Fe magnetic moments with $MoS_2$ through *d*-orbital hybridization. In this context, exploring the valley Zeeman splitting effect in the recently discovered room-temperature ferromagnetic V-doped TMD monolayers [56] would also be very interesting.

## 4. Concluding Remarks and Outlooks

This article has provided a critical review of exchange bias and related emerging phenomena over a wide range of heterostructure systems possessing magnetically coupled vdW/vdW and vdW/non-vdW interfaces. We show that while the EB effects observed in these heterostructures are sizable and interesting for spintronics applications, their origins and association with other effects such as MP and charge transfer are unclear and deserve further studies. $Fe_3GeTe_2$ is emerging as an excellent candidate for exploring 2D van der Waals spintronics. However, this material can be oxidized easily when exposed to air thereby forming a surface layer that possesses a different magnetic order relative to its interior FM layer. While some have suggested that the formation of AFM ordering in this oxidized layer and its interface with the interior FM layer induce the EB effect, there is no solid evidence for the existence of AFM ordering and the resultant FM/AFM coupling. The situation becomes more complex when the EB effect has been reported in $Fe_3GeTe_2$-based heterostructures with various magnetic interfaces including $Fe_3GeTe_2/CrCl_3$ [16], $Fe_3GeTe_2/FePS_2$ [62], and $Fe_3GeTe_2/MnPS_3$ and $Fe_3GeTe_2/MnPSe_3$ [18]. Due to the uncontrolled oxidization effect and the undefined configuration of spins within the surface layer of $Fe_3GeTe_2$, the EB mechanism in these heterostructures cannot be simply attributed to FM/AFM coupling. Further studies on the effect of oxidization on the magnetism and EB in $Fe_3GeTe_2$ and its heterostructures are crucial. Other effects such as magnetic proximity caused by the FM layer ($Fe_3GeTe_2$) on the AFM spins of the adjacent layer (e.g., $CrCl_3$) at the FM/AFM interface and hence on the EB behavior have remained an open question. Since $Fe_3GeTe_2$ is metallic and $CrCl_3$ is semiconducting, there possibly exists "charge transfer" that occurs across this metal/semiconductor interface mediating the interfacial magnetism in this system, which should be investigated further. Given the fact that the EB effect has been observed in $VSe_2/MoS_2$ and $Fe/MoS_2$, it is reasonable to argue that the FM/AFM interface is not a necessary condition for achieving the EB effect in vdW magnetic heterostructures. Instead, interfacing a FM vdW material with a spin cluster or a SG vdW material could also result in a sizable EB effect [154]. In most cases, however, the large EB effect was observed at low temperature (below 100 K). Therefore, there is a pressing need for creating new heterostructure systems that can bring the large EB effect to room temperature. It is also necessary to develop new theoretical models that can better interpret the EB and associated effects in such heterostructures.

From a materials synthesis perspective, it is very challenging to obtain a clean and surface oxidization-free 2D $Fe_3GeTe_2$ from mechanical exfoliation of its single crystal. Consequently, the effect of oxidization on the magnetism and EB in $Fe_3GeTe_2$ and its heterostructures cannot be quantified. A possible solution to this problem is to employ molecular beam epitaxy (MBE) to grow clean $Fe_3GeTe_2$ films in high vacuum and cap the surfaces of the films with thin non-magnetic capping layers (e.g., Ta) to prevent or minimize the films from oxidization. Using this approach, it is also possible to create clean $Fe_3GeTe_2$-based heterostructures that would enable enhanced SV, SOT, ANE, and SSE effects. Nonetheless, the effect of the substrate on the magnetism of $Fe_3GeTe_2$ films should be taken into careful consideration [73][74].

Since the magnetism of $Fe_3GeTe_2$ can be controlled by external stimuli such as ionic gating [76] or light [135,155], it will be extremely interesting to explore electrical and/or optical manipulation of these effects to harness new applications in spintronics, opto-spintronics, spin-caloritronics, opto-spin-caloritronics, valleytronics, and opto-valleytronics. While $Fe_3GeTe_2$ orders ferromagnetically well below 300 K, its Fe-rich compounds (e.g., $Fe_5GeTe_2$) have recently been reported to show Curie temperatures near room temperature [53]. Future research may thus be focused on $Fe_5GeTe_2$ and its heterostructures, which have potential to fulfil the requirements of spintronic devices operating at ambient temperatures. Introducing the 2D magnetism into a TMD semiconducting monolayer by employing the magnetic proximity effect of an adjacent magnetic layer with high $T_C$ such as a $Fe_5GeTe_2$ film may provide a novel strategy to achieve long-range magnetic interactions and boost the Curie temperature of TMDs to above room temperature without adding structural disorder to the TMD lattice. The exploration of SOT switching of 2D magnetic materials such as $Fe_5GeTe_2$ films with high Curie temperature ($T_C > 300$ K) will also benefit the application of SOT-based spintronic devices. Novel $Fe_5GeTe_2$-based heterostructures such as TMD/$Fe_5GeTe_2$ (TMD = $MoS_2$, $WS_2$, $WSe_2$) can be designed and fabricated using the same MBE technique, and the MP effect induced by the $Fe_5GeTe_2$ layer on the TMD layer can be exploited to tune the spin and valley-polarization phenomena at room temperature [120][125][156][157]. The recent observation of the hysteretic magneto response of the exciton emission of quantum emitters in monolayer $WSe_2$ interfaced with $Fe_3GeTe_2$ has indeed demonstrated a new degree of freedom for spin and $g$-factor manipulation of quantum states [158].

Finally, we note that the magnetism of 2D magnetic materials such as $Fe_3GeTe_2$ and V-doped TMDs is very sensitive to strain [159]. Therefore, it is possible to use mechanical strain to manipulate the exchange magnetic anisotropy field and spin dynamics in these systems if they are transferred or grown on flexible substrates [160]. This would open a new research direction in the field of 2D vdW straintronics.

## Acknowledgments

Research was supported by the U.S. Department of Energy, Office of Basic Energy Sciences, Division of Materials Sciences and Engineering under Award No. DE-FG02-07ER 46438. The authors are grateful to Prof. Hari Srikanth of the University of South Florida and Prof. Mingzhong Wu of Colorado State University for their useful comments and suggestions.

# References

[1]    F. Hellman, A. Hoffmann, Y. Tserkovnyak, G.S.D. Beach, E.E. Fullerton, C. Leighton, A.H. MacDonald, D.C. Ralph, D.A. Arena, H.A. Dürr, others, Interface-induced phenomena in magnetism, Rev. Mod. Phys. 89 (2017) 25006.

[2]    M. Gibertini, M. Koperski, A.F. Morpurgo, K.S. Novoselov, Magnetic 2D materials and heterostructures, Nat. Nanotechnol. 14 (2019) 408–419.

[3]    J.F. Sierra, J. Fabian, R.K. Kawakami, S. Roche, S.O. Valenzuela, Van der Waals heterostructures for spintronics and opto-spintronics, Nat. Nanotechnol. 16 (2021) 856–868.

[4]    E.C. Ahn, 2D materials for spintronic devices, Npj 2D Mater. Appl. 4 (2020) 1–14.

[5]    W. Tang, H. Liu, Z. Li, A. Pan, Y.-J. Zeng, Spin-Orbit Torque in Van der Waals-Layered Materials and Heterostructures, Adv. Sci. 8 (2021) 2100847.

[6]    Y. Liu, Q. Shao, Two-dimensional materials for energy-efficient spin--orbit torque devices, ACS Nano. 14 (2020) 9389–9407.

[7]    H.Y. Hwang, Y. Iwasa, M. Kawasaki, B. Keimer, N. Nagaosa, Y. Tokura, Emergent phenomena at oxide interfaces, Nat. Mater. 11 (2012) 103–113.

[8]    A. Bhattacharya, S.J. May, Magnetic oxide heterostructures, Annu. Rev. Mater. Res. 44 (2014) 65–90.

[9]    J. Nogués, I.K. Schuller, Exchange bias, J. Magn. Magn. Mater. 192 (1999) 203–232.

[10]   J. Nogués, J. Sort, V. Langlais, V. Skumryev, S. Suriñach, J.S. Muñoz, M.D. Baró, Exchange bias in nanostructures, Phys. Rep. 422 (2005) 65–117.

[11]   K. O'grady, L.E. Fernandez-Outon, G. Vallejo-Fernandez, A new paradigm for exchange bias in polycrystalline thin films, J. Magn. Magn. Mater. 322 (2010) 883–899.

[12]   S. Giri, M. Patra, S. Majumdar, Exchange bias effect in alloys and compounds, J. Phys. Condens. Matter. 23 (2011) 73201.

[13]   M.-H. Phan, J. Alonso, H. Khurshid, P. Lampen-Kelley, S. Chandra, K. Stojak Repa, Z. Nemati, R. Das, Ó. Iglesias, H. Srikanth, Exchange bias effects in iron oxide-based nanoparticle systems, Nanomaterials. 6 (2016) 221.

[14]   P.K. Manna, S.M. Yusuf, Two interface effects: Exchange bias and magnetic proximity, Phys. Rep. 535 (2014) 61–99.

[15]   T. Blachowicz, A. Ehrmann, Exchange bias in thin films—An update, Coatings. 11 (2021) 122.

[16]   R. Zhu, W. Zhang, W. Shen, P.K.J. Wong, Q. Wang, Q. Liang, Z. Tian, Y. Zhai, C. Qiu, A.T.S. Wee, Exchange Bias in van der Waals $CrCl_3/Fe_3GeTe_2$ Heterostructures, Nano Lett. 20 (2020) 5030–5035.

[17] H. Dai, H. Cheng, M. Cai, Q. Hao, Y. Xing, H. Chen, X. Chen, X. Wang, J.-B. Han, Enhancement of the Coercive Field and Exchange Bias Effect in $Fe_3GeTe_2/MnPX_3$ (X= S and Se) van der Waals Heterostructures, ACS Appl. Mater. & Interfaces. (2021).

[18] G. Hu, Y. Zhu, J. Xiang, T.-Y. Yang, M. Huang, Z. Wang, Z. Wang, P. Liu, Y. Zhang, C. Feng, others, Antisymmetric Magnetoresistance in a van der Waals Antiferromagnetic/Ferromagnetic Layered $MnPS_3/Fe_3GeTe_2$ Stacking Heterostructure, ACS Nano. 14 (2020) 12037–12044.

[19] B. Huang, M.A. McGuire, A.F. May, D. Xiao, P. Jarillo-Herrero, X. Xu, Emergent phenomena and proximity effects in two-dimensional magnets and heterostructures, Nat. Mater. 19 (2020) 1276–1289.

[20] W. Li, Y. Zeng, Z. Zhao, B. Zhang, J. Xu, X. Huang, Y. Hou, 2D Magnetic Heterostructures and Their Interface Modulated Magnetism, ACS Appl. Mater. & Interfaces. 13 (2021) 50591–50601.

[21] Y. Yao, X. Zhan, M.G. Sendeku, P. Yu, F.T. Dajan, N. Li, J. Wang, C. Zhu, F. Wang, Z. Wang, others, Recent progress on emergent two-dimensional magnets and heterostructures, Nanotechnology. (2021).

[22] G. Hu, B. Xiang, Recent Advances in Two-Dimensional Spintronics, Nanoscale Res. Lett. 15 (2020) 1–17.

[23] W.H. Meiklejohn, C.P. Bean, New magnetic anisotropy, Phys. Rev. 102 (1956) 1413.

[24] K. Li, Y. Wu, Z. Guo, Y. Zheng, G. Han, J. Qiu, P. Luo, L. An, T. Zhou, Exchange coupling and its applications in magnetic data storage, J. Nanosci. Nanotechnol. 7 (2007) 13–45.

[25] J.C.S. Kools, Exchange-biased spin-valves for magnetic storage, IEEE Trans. Magn. 32 (1996) 3165–3184.

[26] S.S.P. Parkin, K.P. Roche, M.G. Samant, P.M. Rice, R.B. Beyers, R.E. Scheuerlein, E.J. O'sullivan, S.L. Brown, J. Bucchigano, D.W. Abraham, others, Exchange-biased magnetic tunnel junctions and application to nonvolatile magnetic random access memory, J. Appl. Phys. 85 (1999) 5828–5833.

[27] J.R. Childress, M.J. Carey, R.J. Wilson, N. Smith, C. Tsang, M.K. Ho, K. Carey, S.A. MacDonald, L.M. Ingall, B.A. Gurney, IrMn spin-valves for high density recording, IEEE Trans. Magn. 37 (2001) 1745–1748.

[28] H. Ohldag, A. Scholl, F. Nolting, E. Arenholz, S. Maat, A.T. Young, M. Carey, J. Stöhr, Correlation between exchange bias and pinned interfacial spins, Phys. Rev. Lett. 91 (2003) 17203.

[29] I.K. Schuller, R. Morales, X. Batlle, U. Nowak, G. Güntherodt, Role of the antiferromagnetic bulk spins in exchange bias, J. Magn. Magn. Mater. 416 (2016) 2–9.

[30] R. Carpenter, G. Vallejo-Fernandez, K. O'Grady, Interfacial spin cluster effects in exchange bias systems, J. Appl. Phys. 115 (2014) 17D715.

[31]  A.E. Berkowitz, J.-I. Hong, S.K. McCall, E. Shipton, K.T. Chan, T. Leo, D.J. Smith, Refining the exchange anisotropy paradigm: Magnetic and microstructural heterogeneity at the Permalloy-CoO interface, Phys. Rev. B. 81 (2010) 134404.

[32]  M. Ali, P. Adie, C.H. Marrows, D. Greig, B.J. Hickey, R.L. Stamps, Exchange bias using a spin glass, Nat. Mater. 6 (2007) 70–75.

[33]  G. Zhou, X. Guan, Y. Bai, Z. Quan, F. Jiang, X. Xu, Interfacial spin glass state and exchange bias in the epitaxial La 0.7 Sr 0.3 MnO 3/LaNiO 3 bilayer, Nanoscale Res. Lett. 12 (2017) 1–6.

[34]  E. Maniv, R.A. Murphy, S.C. Haley, S. Doyle, C. John, A. Maniv, S.K. Ramakrishna, Y.-L. Tang, P. Ercius, R. Ramesh, others, Exchange bias due to coupling between coexisting antiferromagnetic and spin-glass orders, Nat. Phys. 17 (2021) 525–530.

[35]  E. Lachman, R.A. Murphy, N. Maksimovic, R. Kealhofer, S. Haley, R.D. McDonald, J.R. Long, J.G. Analytis, Exchange biased anomalous Hall effect driven by frustration in a magnetic kagome lattice, Nat. Commun. 11 (2020) 1–8.

[36]  R.A. Murphy, L.E. Darago, M.E. Ziebel, E.A. Peterson, E.W. Zaia, M.W. Mara, D. Lussier, E.O. Velasquez, D.K. Shuh, J.J. Urban, others, Exchange Bias in a Layered Metal--Organic Topological Spin Glass, ACS Cent. Sci. 7 (2021) 1317–1326.

[37]  I. Žutić, A. Matos-Abiague, B. Scharf, H. Dery, K. Belashchenko, Proximitized materials, Mater. Today. 22 (2019) 85–107.

[38]  F. Magnus, M.E. Brooks-Bartlett, R. Moubah, R.A. Procter, G. Andersson, T.P.A. Hase, S.T. Banks, B. Hjörvarsson, Long-range magnetic interactions and proximity effects in an amorphous exchange-spring magnet, Nat. Commun. 7 (2016) 1–7.

[39]  M. Li, W. Cui, J. Yu, Z. Dai, Z. Wang, F. Katmis, W. Guo, J. Moodera, Magnetic proximity effect and interlayer exchange coupling of ferromagnetic/topological insulator/ferromagnetic trilayer, Phys. Rev. B. 91 (2015) 14427.

[40]  I. Vobornik, U. Manju, J. Fujii, F. Borgatti, P. Torelli, D. Krizmancic, Y.S. Hor, R.J. Cava, G. Panaccione, Magnetic proximity effect as a pathway to spintronic applications of topological insulators, Nano Lett. 11 (2011) 4079–4082.

[41]  D. Zhong, K.L. Seyler, X. Linpeng, N.P. Wilson, T. Taniguchi, K. Watanabe, M.A. McGuire, K.-M.C. Fu, D. Xiao, W. Yao, others, Layer-resolved magnetic proximity effect in van der Waals heterostructures, Nat. Nanotechnol. 15 (2020) 187–191.

[42]  C. Tang, Z. Zhang, S. Lai, Q. Tan, W. Gao, Magnetic proximity effect in graphene/CrBr3 van der Waals heterostructures, Adv. Mater. 32 (2020) 1908498.

[43]  S.M. Suturin, V. V Fedorov, A.G. Banshchikov, D.A. Baranov, K. V Koshmak, P. Torelli, J. Fujii, G. Panaccione, K. Amemiya, M. Sakamaki, others, Proximity effects and exchange bias in Co/MnF$_2$ (111) heterostructures studied by x-ray magnetic circular dichroism, J. Phys. Condens. Matter. 25 (2012) 46002.

[44] J. Van Lierop, K.-W. Lin, J.-Y. Guo, H. Ouyang, B.W. Southern, Proximity effects in an exchange-biased $Ni_{80}Fe_{20}/Co_3O_4$ thin film, Phys. Rev. B. 75 (2007) 134409.

[45] B. Huang, G. Clark, E. Navarro-Moratalla, D.R. Klein, R. Cheng, K.L. Seyler, D. Zhong, E. Schmidgall, M.A. McGuire, D.H. Cobden, others, Layer-dependent ferromagnetism in a van der Waals crystal down to the monolayer limit, Nature. 546 (2017) 270–273.

[46] C. Gong, L. Li, Z. Li, H. Ji, A. Stern, Y. Xia, T. Cao, W. Bao, C. Wang, Y. Wang, others, Discovery of intrinsic ferromagnetism in two-dimensional van der Waals crystals, Nature. 546 (2017) 265–269.

[47] R. Das, J.A. Cardarelli, M.H. Phan, H. Srikanth, Magnetically tunable iron oxide nanotubes for multifunctional biomedical applications, J. Alloys Compd. 789 (2019) 323–329. https://doi.org/10.1016/J.JALLCOM.2019.03.024.

[48] J.R. Schaibley, H. Yu, G. Clark, P. Rivera, J.S. Ross, K.L. Seyler, W. Yao, X. Xu, Valleytronics in 2D materials, Nat. Rev. Mater. 1 (2016) 1–15.

[49] M. Bonilla, S. Kolekar, Y. Ma, H.C. Diaz, V. Kalappattil, R. Das, T. Eggers, H.R. Gutierrez, M.-H. Phan, M. Batzill, Strong room-temperature ferromagnetism in VSe 2 monolayers on van der Waals substrates, Nat. Nanotechnol. 13 (2018) 289–293.

[50] D.J. O'Hara, T. Zhu, A.H. Trout, A.S. Ahmed, Y.K. Luo, C.H. Lee, M.R. Brenner, S. Rajan, J.A. Gupta, D.W. McComb, others, Room temperature intrinsic ferromagnetism in epitaxial manganese selenide films in the monolayer limit, Nano Lett. 18 (2018) 3125–3131.

[51] R. Chua, J. Zhou, X. Yu, W. Yu, J. Gou, R. Zhu, L. Zhang, M. Liu, M.B.H. Breese, W. Chen, others, Room temperature ferromagnetism of monolayer chromium telluride with perpendicular magnetic anisotropy, Adv. Mater. 33 (2021) 2103360.

[52] Z. Fei, B. Huang, P. Malinowski, W. Wang, T. Song, J. Sanchez, W. Yao, D. Xiao, X. Zhu, A.F. May, others, Two-dimensional itinerant ferromagnetism in atomically thin Fe 3 GeTe 2, Nat. Mater. 17 (2018) 778–782.

[53] A.F. May, D. Ovchinnikov, Q. Zheng, R. Hermann, S. Calder, B. Huang, Z. Fei, Y. Liu, X. Xu, M.A. McGuire, Ferromagnetism near room temperature in the cleavable van der Waals crystal $Fe_5GeTe_2$, ACS Nano. 13 (2019) 4436–4442.

[54] J.-U. Lee, S. Lee, J.H. Ryoo, S. Kang, T.Y. Kim, P. Kim, C.-H. Park, J.-G. Park, H. Cheong, Ising-type magnetic ordering in atomically thin $FePS_3$, Nano Lett. 16 (2016) 7433–7438.

[55] S.J. Yun, D.L. Duong, D.M. Ha, K. Singh, T.L. Phan, W. Choi, Y.-M. Kim, Y.H. Lee, Ferromagnetic order at room temperature in monolayer $WSe_2$ semiconductor via vanadium dopant, Adv. Sci. 7 (2020) 1903076.

[56] Y.T.H. Pham, M. Liu, V.O. Jimenez, Z. Yu, V. Kalappattil, F. Zhang, K. Wang, T. Williams, M. Terrones, M.-H. Phan, Tunable ferromagnetism and thermally induced spin flip in Vanadium-doped tungsten diselenide monolayers at room temperature, Adv. Mater. 32 (2020) 2003607.

[57] F. Zhang, B. Zheng, A. Sebastian, D.H. Olson, M. Liu, K. Fujisawa, Y.T.H. Pham, V.O. Jimenez, V. Kalappattil, L. Miao, others, Monolayer vanadium-doped tungsten disulfide: a room-temperature dilute magnetic semiconductor, Adv. Sci. 7 (2020) 2001174.

[58] S. Fu, K. Kang, K. Shayan, A. Yoshimura, S. Dadras, X. Wang, L. Zhang, S. Chen, N. Liu, A. Jindal, others, Enabling room temperature ferromagnetism in monolayer $MoS_2$ via in situ iron-doping, Nat. Commun. 11 (2020) 1–8.

[59] M.-H. Phan, M.T. Trinh, T. Eggers, V. Kalappattil, K. Uchida, L.M. Woods, M. Terrones, A perspective on two-dimensional van der Waals opto-spin-caloritronics, Appl. Phys. Lett. 119 (2021) 250501. https://doi.org/10.1063/5.0069088.

[60] Y.L. Huang, W. Chen, A.T.S. Wee, Two-dimensional magnetic transition metal chalcogenides, SmartMat. 2 (2021) 139-153.

[61] H.K. Gweon, S.Y. Lee, H.Y. Kwon, J. Jeong, H.J. Chang, K.-W. Kim, Z.Q. Qiu, H. Ryu, C. Jang, J.W. Choi, Exchange Bias in Weakly Interlayer-Coupled van der Waals Magnet $Fe_3GeTe_2$, Nano Lett. 21 (2021) 1672–1678.

[62] L. Zhang, X. Huang, H. Dai, M. Wang, H. Cheng, L. Tong, Z. Li, X. Han, X. Wang, L. Ye, others, Proximity-Coupling-Induced Significant Enhancement of Coercive Field and Curie Temperature in 2D van der Waals Heterostructures, Adv. Mater. 32 (2020) 2002032.

[63] Y. Zhang, H. Xu, C. Yi, X. Wang, Y. Huang, J. Tang, J. Jiang, C. He, M. Zhao, T. Ma, others, Exchange bias and spin--orbit torque in the $Fe_3GeTe_2$-based heterostructures prepared by vacuum exfoliation approach, Appl. Phys. Lett. 118 (2021) 262406.

[64] V. Kalappattil et al., Unpublished work.

[65] S. Bhattacharya, S.K. Saha, Design of 2D Ferromagnets With High Exchange Bias in Fe-$MoS_2$ Heterostructure For Spintronic Applications, in: Macromol. Symp., 2017: p. 1600183.

[66] C. Jiang, Y. Wang, Y. Zhang, H. Wang, Q. Chen, J. Wan, Robust Half-Metallic Magnetism in Two-Dimensional $Fe/MoS_2$, J. Phys. Chem. C. 122 (2018) 21617–21622.

[67] Q. Chen, J. Liang, B. Fang, Y. Zhu, J. Wang, W. Lv, W. Lv, J. Cai, Z. Huang, Y. Zhai, others, Proximity effect of a two-dimensional van der Waals magnet $Fe_3GeTe_2$ on nickel films, Nanoscale. 13 (2021) 14688–14693.

[68] H. Idzuchi, A.E. Llacsahuanga Allcca, X.C. Pan, K. Tanigaki, Y.P. Chen, Increased Curie temperature and enhanced perpendicular magneto anisotropy of $Cr_2Ge_2Te_6/NiO$ heterostructures, Appl. Phys. Lett. 115 (2019) 232403.

[69] N. Liu, S. Zhou, J. Zhao, High-Curie-temperature ferromagnetism in bilayer $CrI_3$ on bulk semiconducting substrates, Phys. Rev. Mater. 4 (2020) 94003.

[70] D.S. Kim, J.Y. Kee, J.-E. Lee, Y. Liu, Y. Kim, N. Kim, C. Hwang, W. Kim, C. Petrovic, D.R. Lee, others, Surface oxidation in a van der Waals ferromagnet $Fe_{3-x}GeTe_2$, Curr. Appl. Phys. 30 (2021) 40-45.

[71] L. Peng, F.S. Yasin, T.-E. Park, S.J. Kim, X. Zhang, T. Nagai, K. Kimoto, S. Woo, X. Yu, Tunable Neel-Bloch magnetic twists in $Fe_3GeTe_2$ with van der Waals structure, Adv. Funct. Mater. 31 (2021) 2103583.

[72] Y. Wu, S. Zhang, J. Zhang, W. Wang, Y.L. Zhu, J. Hu, G. Yin, K. Wong, C. Fang, C. Wan, others, Néel-type skyrmion in $WTe_2/Fe_3GeTe_2$ van der Waals heterostructure, Nat. Commun. 11 (2020) 1–6.

[73] L. Zhang, L. Song, H. Dai, J.-H. Yuan, M. Wang, X. Huang, L. Qiao, H. Cheng, X. Wang, W. Ren, others, Substrate-modulated ferromagnetism of two-dimensional $Fe_3GeTe_2$, Appl. Phys. Lett. 116 (2020) 42402.

[74] J.M.J. Lopes, D. Czubak, E. Zallo, A.I. Figueroa, C. Guillemard, M. Valvidares, J.R. Zuazo, J. López-Sanchéz, S.O. Valenzuela, M. Hanke, others, Large-area van der Waals epitaxy and magnetic characterization of $Fe_3GeTe_2$ films on graphene, 2D Mater. 8 (2021) 041001. (2021).

[75] S. Wei, X. Liao, C. Wang, J. Li, H. Zhang, Y.-J. Zeng, J. Linghu, H. Jin, Y. Wei, Emerging intrinsic magnetism in two-dimensional materials: theory and applications, 2D Mater. 8 (2020) 12005.

[76] Y. Deng, Y. Yu, Y. Song, J. Zhang, N.Z. Wang, Z. Sun, Y. Yi, Y.Z. Wu, S. Wu, J. Zhu, others, Gate-tunable room-temperature ferromagnetism in two-dimensional $Fe_3GeTe_2$, Nature. 563 (2018) 94–99.

[77] G. Zheng, W.-Q. Xie, S. Albarakati, M. Algarni, C. Tan, Y. Wang, J. Peng, J. Partridge, L. Farrar, J. Yi, others, Gate-tuned interlayer coupling in van der Waals ferromagnet $Fe_3GeTe_2$ nanoflakes, Phys. Rev. Lett. 125 (2020) 47202.

[78] K. Pei, S. Liu, E. Zhang, X. Zhao, L. Yang, L. Ai, Z. Li, F. Xiu, R. Che, Anomalous Spin Behavior in $Fe_3GeTe_2$ Driven by Current Pulses, ACS Nano. 14 (2020) 9512–9520.

[79] R. Fujimura, R. Yoshimi, M. Mogi, A. Tsukazaki, M. Kawamura, K.S. Takahashi, M. Kawasaki, Y. Tokura, Current-induced magnetization switching at charge-transferred interface between topological insulator $(Bi, Sb)_2Te_3$ and van der Waals ferromagnet $Fe_3GeTe_2$, Appl. Phys. Lett. 119 (2021) 32402.

[80] S. Liu, K. Yang, W. Liu, E. Zhang, Z. Li, X. Zhang, Z. Liao, W. Zhang, J. Sun, Y. Yang, others, Two-dimensional ferromagnetic superlattices, Natl. Sci. Rev. 7 (2020) 745–754.

[81] X. Lin, J. Ni, Layer-dependent intrinsic anomalous Hall effect in $Fe_3GeTe_2$, Phys. Rev. B. 100 (2019) 85403.

[82] D. Kim, S. Park, J. Lee, J. Yoon, S. Joo, T. Kim, K. Min, S.-Y. Park, C. Kim, K.-W. Moon, others, Antiferromagnetic coupling of van der Waals ferromagnetic $Fe_3GeTe_2$, Nanotechnology. 30 (2019) 245701.

[83] S.J. Kim, D. Choi, K.-W. Kim, K.-Y. Lee, D.-H. Kim, S. Hong, J. Suh, C. Lee, S.K. Kim, T.-E. Park, others, Interface Engineering of Magnetic Anisotropy in van der Waals Ferromagnet-based Heterostructures, ACS Nano. 15 (2021) 16395–16403.

[84] X. Cai, T. Song, N.P. Wilson, G. Clark, M. He, X. Zhang, T. Taniguchi, K. Watanabe, W. Yao, D. Xiao, others, Atomically thin CrCl$_3$: an in-plane layered antiferromagnetic insulator, Nano Lett. 19 (2019) 3993–3998.

[85] K. Kim, S.Y. Lim, J. Kim, J.-U. Lee, S. Lee, P. Kim, K. Park, S. Son, C.-H. Park, J.-G. Park, others, Antiferromagnetic ordering in van der Waals 2D magnetic material MnPS$_3$ probed by Raman spectroscopy, 2D Mater. 6 (2019) 41001.

[86] H. Fu, C.-X. Liu, B. Yan, Exchange bias and quantum anomalous Hall effect in the MnBi$_2$Te$_4$/CrI$_3$ heterostructure, Sci. Adv. 6 (2020) eaaz0948.

[87] Q.I. Yang, A. Kapitulnik, Two-stage proximity-induced gap opening in topological-insulator--insulating-ferromagnet (Bi$_x$Sb$_{1-x}$)$_2$Te$_3$-EuS bilayers, Phys. Rev. B. 98 (2018) 81403.

[88] J. Kim, K.-W. Kim, H. Wang, J. Sinova, R. Wu, Understanding the giant enhancement of exchange interaction in Bi$_2$Se$_3$- EuS Heterostructures, Phys. Rev. Lett. 119 (2017) 27201.

[89] A. Manchon, J. Železn`y, I.M. Miron, T. Jungwirth, J. Sinova, A. Thiaville, K. Garello, P. Gambardella, Current-induced spin-orbit torques in ferromagnetic and antiferromagnetic systems, Rev. Mod. Phys. 91 (2019) 35004.

[90] M. Alghamdi, M. Lohmann, J. Li, P.R. Jothi, Q. Shao, M. Aldosary, T. Su, B.P.T. Fokwa, J. Shi, Highly efficient spin--orbit torque and switching of layered ferromagnet Fe$_3$GeTe$_2$, Nano Lett. 19 (2019) 4400–4405.

[91] H. Zhong, C. Bao, H. Wang, J. Li, Z. Yin, Y. Xu, W. Duan, T.-L. Xia, S. Zhou, Light-Tunable Surface State and Hybridization Gap in Magnetic Topological Insulator MnBi$_8$Te$_{13}$, Nano Lett. 21 (2021) 6080–6086.

[92] S.Z. Butler, S.M. Hollen, L. Cao, Y. Cui, J.A. Gupta, H.R. Gutiérrez, T.F. Heinz, S.S. Hong, J. Huang, A.F. Ismach, others, Progress, challenges, and opportunities in two-dimensional materials beyond graphene, ACS Nano. 7 (2013) 2898–2926.

[93] D.L. Duong, S.J. Yun, Y.H. Lee, van der Waals layered materials: opportunities and challenges, ACS Nano. 11 (2017) 11803–11830.

[94] S.-P. Tsai, C.-Y. Yang, C.-J. Lee, L.-S. Lu, H.-L. Liang, J.-X. Lin, Y.-H. Yu, C.-C. Chen, T.-K. Chung, C.-C. Kaun, others, Room-Temperature Ferromagnetism of Single-Layer MoS$_2$ Induced by Antiferromagnetic Proximity of Yttrium Iron Garnet, Adv. Quantum Technol. 4 (2021) 2000104.

[95] W. Zhang, L. Zhang, P.K.J. Wong, J. Yuan, G. Vinai, P. Torelli, G. van der Laan, Y.P. Feng, A.T.S. Wee, Magnetic transition in monolayer VSe$_2$ via interface hybridization, ACS Nano. 13 (2019) 8997–9004.

[96] G. Vinai, C. Bigi, A. Rajan, M.D. Watson, T.-L. Lee, F. Mazzola, S. Modesti, S. Barua, M.C. Hatnean, G. Balakrishnan, others, Proximity-induced ferromagnetism and chemical reactivity in few-layer VSe 2 heterostructures, Phys. Rev. B. 101 (2020) 35404.

[97] K. Dolui, A. Narayan, I. Rungger, S. Sanvito, Efficient spin injection and giant

magnetoresistance in Fe/MoS$_2$/Fe junctions, Phys. Rev. B. 90 (2014) 41401.

[98]   W. Wang, A. Narayan, L. Tang, K. Dolui, Y. Liu, X. Yuan, Y. Jin, Y. Wu, I. Rungger, S. Sanvito, others, Spin-valve effect in NiFe/MoS$_2$/NiFe junctions, Nano Lett. 15 (2015) 5261–5267.

[99]   H. Lin, F. Yan, C. Hu, Q. Lv, W. Zhu, Z. Wang, Z. Wei, K. Chang, K. Wang, Spin-valve effect in Fe$_3$GeTe$_2$/MoS$_2$/Fe$_3$GeTe$_2$ van der Waals heterostructures, ACS Appl. Mater. & Interfaces. 12 (2020) 43921–43926.

[100]  L. Zhang, T. Li, J. Li, Y. Jiang, J. Yuan, H. Li, Perfect spin filtering effect on Fe$_3$GeTe$_2$-based van der Waals magnetic tunnel junctions, J. Phys. Chem. C. 124 (2020) 27429–27435.

[101]  L. Cai, J. He, Q. Liu, T. Yao, L. Chen, W. Yan, F. Hu, Y. Jiang, Y. Zhao, T. Hu, others, Vacancy-induced ferromagnetism of MoS2 nanosheets, J. Am. Chem. Soc. 137 (2015) 2622–2627.

[102]  N. Feng, W. Mi, Y. Cheng, Z. Guo, U. Schwingenschlögl, H. Bai, Magnetism by interfacial hybridization and p-type doping of MoS$_2$ in Fe$_4$N/MoS$_2$ superlattices: a first-principles study, ACS Appl. Mater. & Interfaces. 6 (2014) 4587–4594.

[103]  V. Zatko, M. Galbiati, S.M.-M. Dubois, M. Och, P. Palczynski, C. Mattevi, P. Brus, O. Bezencenet, M.-B. Martin, B. Servet, others, Band-structure spin-filtering in vertical spin valves based on chemical vapor deposited WS$_2$, ACS Nano. 13 (2019) 14468–14476.

[104]  B. Peng, Q. Li, X. Liang, P. Song, J. Li, K. He, D. Fu, Y. Li, C. Shen, H. Wang, others, Valley polarization of trions and magnetoresistance in heterostructures of MoS$_2$ and yttrium iron garnet, ACS Nano. 11 (2017) 12257–12265.

[105]  W.-Y. Lee, M.-S. Kang, G.-S. Kim, N.-W. Park, K.-Y. Choi, C.T. Le, M.U. Rashid, E. Saitoh, Y.S. Kim, S.-K. Lee, Role of Ferromagnetic Monolayer WSe2 Flakes in the Pt/Y$_3$Fe$_5$O$_{12}$ Bilayer Structure in the Longitudinal Spin Seebeck Effect, ACS Appl. Mater. & Interfaces. 13 (2021) 15783–15790.

[106]  C.-I. Lu, C.-H. Huang, K.-H.O. Yang, K.B. Simbulan, K.-S. Li, F. Li, J. Qi, M. Jugovac, I. Cojocariu, V. Feyer, others, Spontaneously induced magnetic anisotropy in an ultrathin Co/MoS$_2$ heterojunction, Nanoscale Horizons. 5 (2020) 1058–1064.

[107]  G. Dastgeer, M.A. Shehzad, J. Eom, Distinct Detection of Thermally Induced Spin Voltage in Pt/WS$_2$/Ni$_{81}$Fe$_{19}$ by the Inverse Spin Hall Effect, ACS Appl. Mater. & Interfaces. 11 (2019) 48533–48539.

[108]  S. Lee, W. Lee, T. Kikkawa, C.T. Le, M. Kang, G. Kim, A.D. Nguyen, Y.S. Kim, N. Park, E. Saitoh, Enhanced Spin Seebeck Effect in Monolayer Tungsten Diselenide Due to Strong Spin Current Injection at Interface, Adv. Funct. Mater. 30 (2020) 2003192.

[109]  W. Yu, J. Li, T.S. Herng, Z. Wang, X. Zhao, X. Chi, W. Fu, I. Abdelwahab, J. Zhou, J. Dan, others, Chemically exfoliated VSe$_2$ monolayers with room-temperature ferromagnetism, Adv. Mater. 31 (2019) 1903779.

[110] G. Chen, S.T. Howard, A.B. Maghirang III, K.N. Cong, R.A.B. Villaos, L.-Y. Feng, K. Cai, S.C. Ganguli, W. Swiech, E. Morosan, others, Correlating structural, electronic, and magnetic properties of epitaxial $VSe_2$ thin films, Phys. Rev. B. 102 (2020) 115149.

[111] K. Shawulienu, H.M. Nurul, P. Dreher, M. Ilkka, Y. Zhou, S. Jani, M. Rhodri, M.M. Ugeda, K. Hannu-Pekka, P. Liljeroth, others, Electronic and magnetic characterization of epitaxial $VSe_2$ monolayers on superconducting $NbSe_2$, Commun. Phys. 3 (2020).

[112] P.M. Coelho, K. Nguyen Cong, M. Bonilla, S. Kolekar, M.-H. Phan, J. Avila, M.C. Asensio, I.I. Oleynik, M. Batzill, Charge density wave state suppresses ferromagnetic ordering in $VSe_2$ monolayers, J. Phys. Chem. C. 123 (2019) 14089–14096.

[113] R. Chua, J. Yang, X. He, X. Yu, W. Yu, F. Bussolotti, P.K.J. Wong, K.P. Loh, M.B.H. Breese, K.E.J. Goh, others, Can Reconstructed Se-Deficient Line Defects in Monolayer $VSe_2$ Induce Magnetism?, Adv. Mater. 32 (2020) 2000693.

[114] T.J. Kim, S. Ryee, M.J. Han, S. Choi, Dynamical mean-field study of Vanadium diselenide monolayer ferromagnetism, 2D Mater. 7 (2020) 35023.

[115] D.W. Boukhvalov, A. Politano, Unveiling the origin of room-temperature ferromagnetism in monolayer $VSe_2$: the role of extrinsic effects, Nanoscale. 12 (2020) 20875–20882.

[116] Z. Guguchia, A. Kerelsky, D. Edelberg, S. Banerjee, F. von Rohr, D. Scullion, M. Augustin, M. Scully, D.A. Rhodes, Z. Shermadini, others, Magnetism in semiconducting molybdenum dichalcogenides, Sci. Adv. 4 (2018) eaat3672.

[117] Q. Liang, Q. Zhang, X. Zhao, M. Liu, A.T.S. Wee, Defect engineering of two-dimensional transition-metal dichalcogenides: applications, challenges, and opportunities, ACS Nano. 15 (2021) 2165–2181.

[118] X. He, Y. Wang, N. Wu, A.N. Caruso, E. Vescovo, K.D. Belashchenko, P.A. Dowben, C. Binek, Robust isothermal electric control of exchange bias at room temperature, Nat. Mater. 9 (2010) 579–585.

[119] Q.L. He, X. Kou, A.J. Grutter, G. Yin, L. Pan, X. Che, Y. Liu, T. Nie, B. Zhang, S.M. Disseler, others, Tailoring exchange couplings in magnetic topological-insulator/antiferromagnet heterostructures, Nat. Mater. 16 (2017) 94–100.

[120] D. Soriano, J.L. Lado, Exchange-bias controlled correlations in magnetically encapsulated twisted van der Waals dichalcogenides, J. Phys. D. Appl. Phys. 53 (2020) 474001.

[121] K. Zhang, S. Han, Y. Lee, M.J. Coak, J. Kim, I. Hwang, S. Son, J. Shin, M. Lim, D. Jo, others, Gigantic Current Control of Coercive Field and Magnetic Memory Based on Nanometer-Thin Ferromagnetic van der Waals $Fe_3GeTe_2$, Adv. Mater. 33 (2021) 2004110.

[122] M. Li, C.-Z. Chang, B.J. Kirby, M.E. Jamer, W. Cui, L. Wu, P. Wei, Y. Zhu, D. Heiman, J. Li, others, Proximity-driven enhanced magnetic order at ferromagnetic-insulator-magnetic-topological-insulator interface, Phys. Rev. Lett. 115 (2015) 87201.

[123] T. Song, X. Cai, M.W.-Y. Tu, X. Zhang, B. Huang, N.P. Wilson, K.L. Seyler, L. Zhu, T.

Taniguchi, K. Watanabe, others, Giant tunneling magnetoresistance in spin-filter van der Waals heterostructures, Science 360 (2018) 1214–1218.

[124] J. Zhou, J. Qiao, C.-G. Duan, A. Bournel, K.L. Wang, W. Zhao, Large tunneling magnetoresistance in $VSe_2/MoS_2$ magnetic tunnel junction, ACS Appl. Mater. & Interfaces. 11 (2019) 17647–17653.

[125] C. Hu, F. Yan, Y. Li, K. Wang, Vertical $WS_2$ Spin Valve with Ohmic Property Based on $Fe_3GeTe_2$ Electrodes, Chinese Phys. B. 30 (2021) 097505.

[126] W. Zhu, H. Lin, F. Yan, C. Hu, Z. Wang, L. Zhao, Y. Deng, Z.R. Kudrynskyi, T. Zhou, Z.D. Kovalyuk, others, Large Tunneling Magnetoresistance in van der Waals Ferromagnet/Semiconductor Heterojunctions, Adv. Mater. 33 (2021) 2104658.

[127] Z. Wang, D. Sapkota, T. Taniguchi, K. Watanabe, D. Mandrus, A.F. Morpurgo, Tunneling spin valves based on $Fe_3GeTe_2/hBN/Fe_3GeTe_2$ van der Waals heterostructures, Nano Lett. 18 (2018) 4303–4308.

[128] S. Albarakati, C. Tan, Z.-J. Chen, J.G. Partridge, G. Zheng, L. Farrar, E.L.H. Mayes, M.R. Field, C. Lee, Y. Wang, others, Antisymmetric magnetoresistance in van der Waals $Fe_3GeTe_2/graphite/Fe_3GeTe_2$ trilayer heterostructures, Sci. Adv. 5 (2019) eaaw0409.

[129] X. Wang, J. Tang, X. Xia, C. He, J. Zhang, Y. Liu, C. Wan, C. Fang, C. Guo, W. Yang, others, Current-driven magnetization switching in a van der Waals ferromagnet $Fe_3GeTe_2$, Sci. Adv. 5 (2019) eaaw8904.

[130] V. Gupta, T.M. Cham, G.M. Stiehl, A. Bose, J.A. Mittelstaedt, K. Kang, S. Jiang, K.F. Mak, J. Shan, R.A. Buhrman, others, Manipulation of the van der Waals Magnet $Cr_2Ge_2Te_6$ by Spin-Orbit Torques, Nano Lett. 20 (2020) 7482–7488.

[131] K. Dolui, M.D. Petrovic, K. Zollner, P. Plechac, J. Fabian, B.K. Nikolic, Proximity spin-orbit torque on a two-dimensional magnet within van der Waals heterostructure: current-driven antiferromagnet-to-ferromagnet reversible nonequilibrium phase transition in bilayer $CrI_3$, Nano Lett. 20 (2020) 2288–2295.

[132] V. Ostwal, T. Shen, J. Appenzeller, Efficient Spin-Orbit Torque Switching of the Semiconducting Van Der Waals Ferromagnet $Cr_2Ge_2Te_6$, Adv. Mater. 32 (2020) 1906021.

[133] H. Xue, M. Tang, Y. Zhang, Z. Ji, X. Qiu, Z. Zhang, Giant Enhancement of Spin-Orbit Torque Efficiency in Pt/Co Bilayers by Inserting a $WSe_2$ under Layer, Adv. Electron. Mater. (2021) 2100684.

[134] Q. Xie, W. Lin, B. Yang, X. Shu, S. Chen, L. Liu, X. Yu, M.B.H. Breese, T. Zhou, M. Yang, others, Giant enhancements of perpendicular magnetic anisotropy and spin-orbit torque by a $MoS_2$ layer, Adv. Mater. 31 (2019) 1900776.

[135] B. Liu, S. Liu, L. Yang, Z. Chen, E. Zhang, Z. Li, J. Wu, X. Ruan, F. Xiu, W. Liu, others, Light-Tunable Ferromagnetism in Atomically Thin $Fe_3GeTe_2$ Driven by Femtosecond Laser Pulse, Phys. Rev. Lett. 125 (2020) 267205.

[136] K. Uchida, S. Takahashi, K. Harii, J. Ieda, W. Koshibae, K. Ando, S. Maekawa, E. Saitoh, Observation of the spin Seebeck effect, Nature. 455 (2008) 778–781.

[137] K.-I. Uchida, Transport phenomena in spin caloritronics, Proc. Japan Acad. Ser. B. 97 (2021) 69–88.

[138] J. Xu, W.A. Phelan, C.-L. Chien, Large anomalous Nernst effect in a van der Waals ferromagnet $Fe_3GeTe_2$, Nano Lett. 19 (2019) 8250–8254.

[139] V. Kalappattil, R. Das, M.-H. Phan, H. Srikanth, Roles of bulk and surface magnetic anisotropy on the longitudinal spin Seebeck effect of Pt/YIG, Sci. Rep. 7 (2017) 13316.

[140] V. Kalappattil, R. Geng, R. Das, M. Pham, H. Luong, T. Nguyen, A. Popescu, L.M. Woods, M. Kläui, H. Srikanth, Giant spin Seebeck effect through an interface organic semiconductor, Mater. Horizons. 7 (2020) 1413–1420.

[141] V. Ortiz Jimenez, Y.T.H. Pham, M. Liu, F. Zhang, Z. Yu, V. Kalappattil, B. Muchharla, T. Eggers, D.L. Duong, M. Terrones, others, Light-Controlled Room Temperature Ferromagnetism in Vanadium-Doped Tungsten Disulfide Semiconducting Monolayers, Adv. Electron. Mater. 7 (2021) 2100030.

[142] S. Zhao, B. Dong, H. Wang, Y. Zhang, H. Wang, Z. Han, H. Zhang, others, Valley manipulation in monolayer transition metal dichalcogenides and their hybrid systems: status and challenges, Reports Prog. Phys. 84 (2021) 026401.

[143] G. Aivazian, Z. Gong, A.M. Jones, R.-L. Chu, J. Yan, D.G. Mandrus, C. Zhang, D. Cobden, W. Yao, X. Xu, Magnetic control of valley pseudospin in monolayer $WSe_2$, Nat. Phys. 11 (2015) 148–152.

[144] D. Xiao, G.-B. Liu, W. Feng, X. Xu, W. Yao, Coupled spin and valley physics in monolayers of $MoS_2$ and other group-VI dichalcogenides, Phys. Rev. Lett. 108 (2012) 196802.

[145] X. Xu, W. Yao, D. Xiao, T.F. Heinz, Spin and pseudospins in layered transition metal dichalcogenides, Nat. Phys. 10 (2014) 343–350.

[146] S. Wu, J.S. Ross, G.-B. Liu, G. Aivazian, A. Jones, Z. Fei, W. Zhu, D. Xiao, W. Yao, D. Cobden, others, Electrical tuning of valley magnetic moment through symmetry control in bilayer $MoS_2$, Nat. Phys. 9 (2013) 149–153.

[147] T. Norden, C. Zhao, P. Zhang, R. Sabirianov, A. Petrou, H. Zeng, Giant valley splitting in monolayer $WS_2$ by magnetic proximity effect, Nat. Commun. 10 (2019) 1–10.

[148] D. Zhong, K.L. Seyler, X. Linpeng, R. Cheng, N. Sivadas, B. Huang, E. Schmidgall, T. Taniguchi, K. Watanabe, M.A. McGuire, others, Van der Waals engineering of ferromagnetic semiconductor heterostructures for spin and valleytronics, Sci. Adv. 3 (2017) e1603113.

[149] K.-A. Min, J. Cha, K. Cho, S. Hong, Ferromagnetic contact between Ni and $MoX_2$ (X= S, Se, or Te) with Fermi-level pinning, 2D Mater. 4 (2017) 24006.

[150] X. Liang, L. Deng, F. Huang, T. Tang, C. Wang, Y. Zhu, J. Qin, Y. Zhang, B. Peng, L. Bi, The magnetic proximity effect and electrical field tunable valley degeneracy in $MoS_2/EuS$ van der Waals heterojunctions, Nanoscale. 9 (2017) 9502–9509.

[151] C. Zhao, T. Norden, P. Zhang, P. Zhao, Y. Cheng, F. Sun, J.P. Parry, P. Taheri, J. Wang, Y. Yang, others, Enhanced valley splitting in monolayer $WSe_2$ due to magnetic exchange field, Nat. Nanotechnol. 12 (2017) 757–762.

[152] K.L. Seyler, D. Zhong, B. Huang, X. Linpeng, N.P. Wilson, T. Taniguchi, K. Watanabe, W. Yao, D. Xiao, M.A. McGuire, others, Valley manipulation by optically tuning the magnetic proximity effect in $WSe_2/CrI_3$ heterostructures, Nano Lett. 18 (2018) 3823–3828.

[153] Q. Li, X. Zhao, L. Deng, Z. Shi, S. Liu, Q. Wei, L. Zhang, Y. Cheng, L. Zhang, H. Lu, others, Enhanced valley Zeeman splitting in Fe-doped monolayer $MoS_2$, ACS Nano. 14 (2020) 4636–4645.

[154] S. Doyle, C. John, E. Maniv, R.A. Murphy, A. Maniv, S.K. Ramakrishna, Y.-L. Tang, R. Ramesh, J.R. Long, A.P. Reyes, others, Tunable giant exchange bias in an intercalated transition metal dichalcogenide, ArXiv Prepr. ArXiv1904.05872. (2019).

[155] J. He, S. Li, A. Bandyopadhyay, T. Frauenheim, Unravelling Photoinduced Interlayer Spin Transfer Dynamics in Two-Dimensional Nonmagnetic-Ferromagnetic van der Waals Heterostructures, Nano Lett. 21 (2021) 3237–3244.

[156] H. Da, Q. Song, P. Dong, H. Ye, X. Yan, Control of spin and valley Hall effects in monolayer transition metal dichalcogenides by magnetic proximity effect, J. Appl. Phys. 127 (2020) 23903.

[157] Y. Tokura, M. Kawasaki, N. Nagaosa, Emergent functions of quantum materials, Nat. Phys. 13 (2017) 1056–1068.

[158] N. Liu, C.M. Gallaro, K. Shayan, A. Mukherjee, B. Kim, J. Hone, N. Vamivakas, S. Strauf, Antiferromagnetic proximity coupling between semiconductor quantum emitters in $WSe_2$ and van der Waals ferromagnets, Nanoscale. 13 (2021) 832–841.

[159] X. Hu, Y. Zhao, X. Shen, A. V Krasheninnikov, Z. Chen, L. Sun, Enhanced ferromagnetism and tunable magnetism in $Fe_3GeTe_2$ monolayer by strain engineering, ACS Appl. Mater. & Interfaces. 12 (2020) 26367–26373.

[160] Z. Zhang, E. Liu, W. Zhang, P.K.J. Wong, Z. Xu, F. Hu, X. Li, J. Tang, A.T.S. Wee, F. Xu, Mechanical Strain manipulation of exchange bias field and spin dynamics in FeCo/IrMn multilayers grown on flexible substrates, ACS Appl. Mater. & Interfaces. 11 (2019) 8258–8265.

**Figure captions**

**Figure 1.** Schematic shows the magnetic proximity (MP) and exchange bias (EB) effects in a van der Waals magnetic heterostructure. $Fe_3GeTe_2$ represents an example of material that preserves a ferromagnetic ordering in the 2D limit and can be interfaced with other material such as $MnPS_3$ to induce both MP and EB effects.

**Figure 2. (c)** Crystal and magnetic structure of $Fe_3GeTe_2$; **(b)** Cross-sectional STEM image of a $Fe_3GeTe_2$ flake capped with an $AlO_x$ (2.5 nm) passivation layer; **(c)** Temperature and cooling-field dependence of the out-of-plane $M-H$ loops measured for the oxidized $Fe_3GeTe_2$ flake ($t = 38$ nm); **(d)** $H_{EB}$ and $H_C$ as functions of $T$ for two different flake thicknesses of $t = 17$ nm (circle) and 38 nm (square). *Reproduced with permission from Gweon et al., Nano Letters 21, 1672 (2021).*

**Figure 3. (a)** Side view of the structure of $CrCl_3$. **(b)** Color plot of normalized $dI_t/dT$ as a function of in-plane magnetic field $H_{in}$ and temperature $T$. AFM: layered antiferromagnetic. PM: paramagnetic. SP: Fully spin polarized. *Reproduced with permission from Cai et al., Nano Letters 19, 3993 (2019);* **(c)** Schematic diagram of the whole device structure. Inset: Optical image of the $CrCl_3$(top)/$Fe_3GeTe_2$ heterostructure on top of the Pt Hall contacts. **(d)** Temperature-dependent magnitudes of the exchange-bias field ($|H_E|$) and coercivity ($H_C$) in the individual $Fe_3GeTe_2$(30 nm) and $CrCl_3$(15 nm, 45 nm)/ $Fe_3GeTe_2$(30 nm) heterostructures. (e) $R_{xy}$-$H$ loops for individual $Fe_3GeTe_2$(30 nm) and $CrCl_3$(15 nm, 45 nm)/$Fe_3GeTe_2$(30 nm) heterostructures, respectively, measured at 2.5 K. *Reproduced with permission from Zhu et al., Nano Letters 20, 5030 (2020).*

**Figure 4. (a)** Magnetic ordering in vdW $Fe_3GeTe_2$ and $FePS_3$ flakes; **(b)** Magnetic susceptibility as a function of temperature for $Fe_3GeTe_2$ (gray curve) and $FePS_3$ (orange curve) single crystals along the c-axis; **(c)** Comparison of the MOKE signals versus the magnetic field at 80 K for $Fe_3GeTe_2$ and FPS/FGT; **(d)** The domain of the $Fe_3GeTe_2$ and $FePS_3$ layers around the interface of $Fe_3GeTe_2$/$FePS_3$

under the different external magnetic fields; **(e)** Comparison of the MOKE signals versus the magnetic field at 80 K for bare $Fe_3GeTe_2$ and the $FePS_3/Fe_3GeTe_2/FePS_3$ heterostructure; **(f)** The magnetic domain of $FePS_3/Fe_3GeTe_2/FePS_3$ at three typical magnetic states. *Reproduced with permission from Zhang et al., Advanced Materials 32, 2002032 (2020).*

**Figure 5. (a)** Crystal structures of the $Fe_3GeTe_2/MnPS_3$ and $Fe_3GeTe_2/MnPSe_3$ heterostructures; **(b)** Polar RMCD signal of $Fe_3GeTe_2$, $Fe_3GeTe_2/MnPS_3$, and $Fe_3GeTe_2/MnPSe_3$ measured at 10 K; **(c)** Temperature dependent coercive fields of the $Fe_3GeTe_2$ flake, $Fe_3GeTe_2/MnPS_3$, and $Fe_3GeTe_2/MnPSe_3$ heterostructures; **(d)** Temperature-dependent exchange bias fields of $Fe_3GeTe_2/MnPS_3$ and $Fe_3GeTe_2/MnPSe_3$ heterostructures. *Reproduced with permission from Dai et al., ACS Appl. Mater. Interfaces 13, 24314 (2021).*

**Figure 6. (a)** Schematics of the $Fe_3GeTe_2/(IrMn$ or $Pt)$ bilayers. Cross-sectional STEM image of the $Fe_3GeTe_2/IrMn$ device fabricated on a $Si/SiO_2$ substrate by the vacuum exfoliation approach. Optical image of the hall-bar device for transport measurements; **(b)** Dependence of $R_{xy}$ on out-of-plane $H$ for IrMn (2 nm)/$Fe_3GeTe_2$ devices (prepared in a high vacuum) measured at 2 K. A clear shift to the left (right) is observed for the field cooling under a field of þ1 (–1 T); **(c)** Hall resistance measured with a current of 66 mA and an external field of 3000 Oe applied in the $x$-$z$ plane and at an angle $b$ away from the $x$ direction. *Reproduced with permission from Zhang et al., Appl. Phys. Lett. 118, 262406 (2021).*

**Figure 7. (a)** Schematic diagram of the $Fe_3GeTe_2/MoS_2/Fe_3GeTe_2$ spin valve encapsulated by a top h-BN capping layer (∼50 nm thick). Junction resistance and MR vs magnetic field at 10 K. ↓ and ↑ denote the out-of-plane magnetizations alignment directions of the $Fe_3GeTe_2$ flakes; *Reproduced with permission from Lin et al. ACS Appl. Mater. Interfaces 12, 43921 (2020).* **(b)** The schematic diagram of the device and magnetotransport setup. An out-of-plane magnetic field was applied to control the magnetization of the two Fe₃GeTe₂ electrodes. Resistance as a function of the perpendicular magnetic field (R–B) of device A at a fixed current bias of 0.1 μA at 10 K; *Reproduced with permission from Zhu et al. Adv. Mater. 2104658 (2021).* **(c)** Sample for Lorentz transmission electron microscopy measurements consisting of 2L WTe₂ and 30L Fe₃GeTe₂. Scale bar: 10 μm. Typical labyrinth domain in 30L Fe₃GeTe₂ thin flakes. Scale bar: 2 μm. From the aligned and stripe-like domain structures of the WTe₂/Fe₃GeTe₂, a Dzyaloshinskii–Moriya interaction energy is estimated to be ~1.0 mJ m$^{-2}$. Scale bar: 2 μm. *Reproduced with permission from Wu et al. Nature Com. 11, 3860 (2020).*

**Figure 8. (a)** Raman spectra of YIG and the MoS₂ single layer on YIG. The Raman spectrum of single-layer MoS₂ reflects the signature E$^1_{2g}$ and A$_{1g}$ modes, which are associated with horizontal and vertical vibration modes as the insets; **(b)** Highlighted MCD spectra of YIG/MoS₂ and the reference MCDs of **(c)** YIG/Al₂O₃/MoS₂ and bare YIG. Arrows in (a) and (b) indicate antiferromagnetically coupled moments in YIG and MoS₂ based on the anti-symmetry of MCD. **(d)** Field-dependent MCD taken with a fixed photon energy at 2.95 and 1.97 eV, corresponding to the optical response of YIG and MoS₂, respectively. **(e)** Exchange between induced moments in MoS₂ assisted without/with BMP. The induced moments locate in the vicinity of sulfur vacancies enable the long-range magnetic interaction through BMP percolation. *Reproduced with permission from Tsai et al. Adv. Quantum Technol. 4, 2000104 (2021).*

**Figure 9. (a)** Schematic shows a bilayer heterostructure composed of monolayer VSe₂ and single crystal MoS₂; **(b)** Magnetic hysteresis loops in FC and ZFC regimes taken at 10 K for the VSe₂/MoS₂ bilayer; **(c)** Temperature dependence of exchange bias field ($H_{EB}$) and coercive field ($H_C$) of the VSe₂/MoS₂ bilayer; **(d)** Cooling field dependence of exchange bias field ($H_{EB}$) for the VSe₂/MoS₂ bilayer.

**Figure 10.** Schematic showing perspective applications (spintronics; opto-spintronics, spin-caloritronics; opto-spin-caloritronics; valleytronics; spin-valleytronics) of van der Waals magnets and their heterostructures in which exchange magnetic anisotropy plays an important role.

**Figure 1**

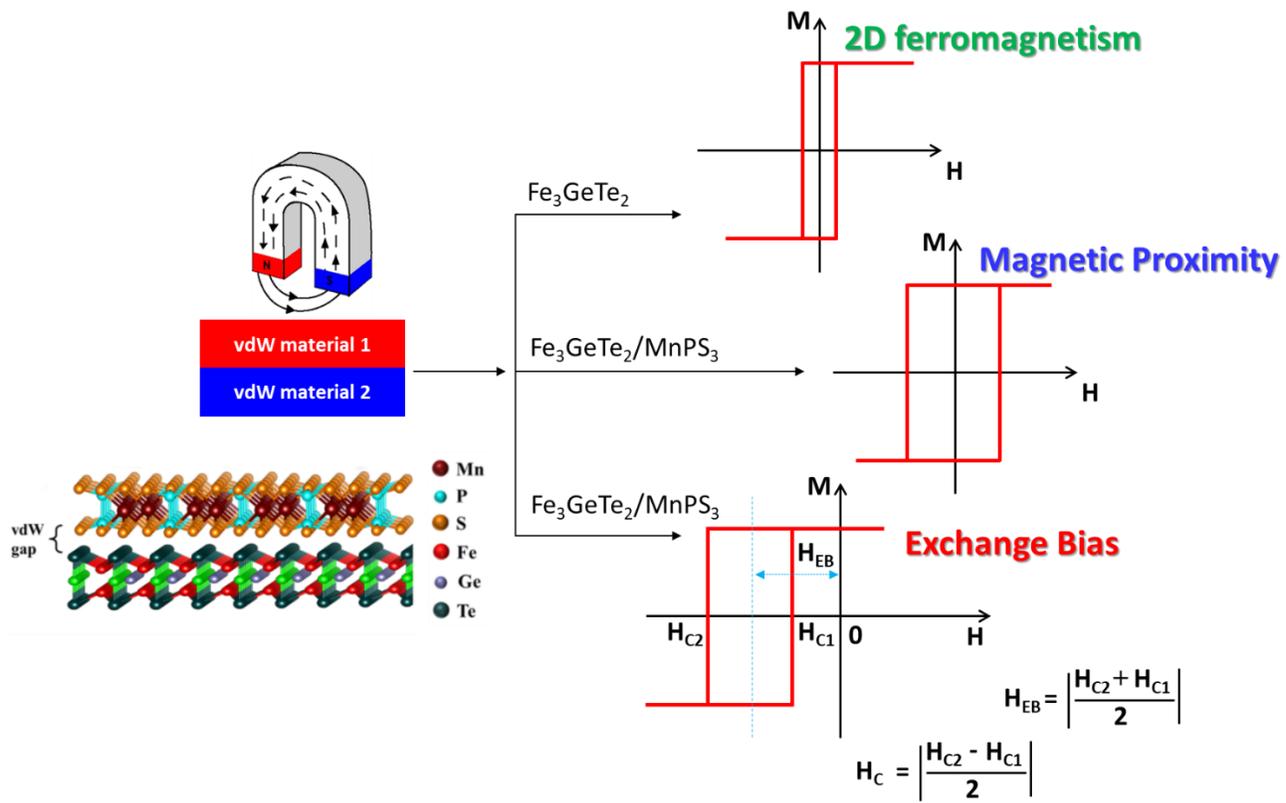

**Figure 2**

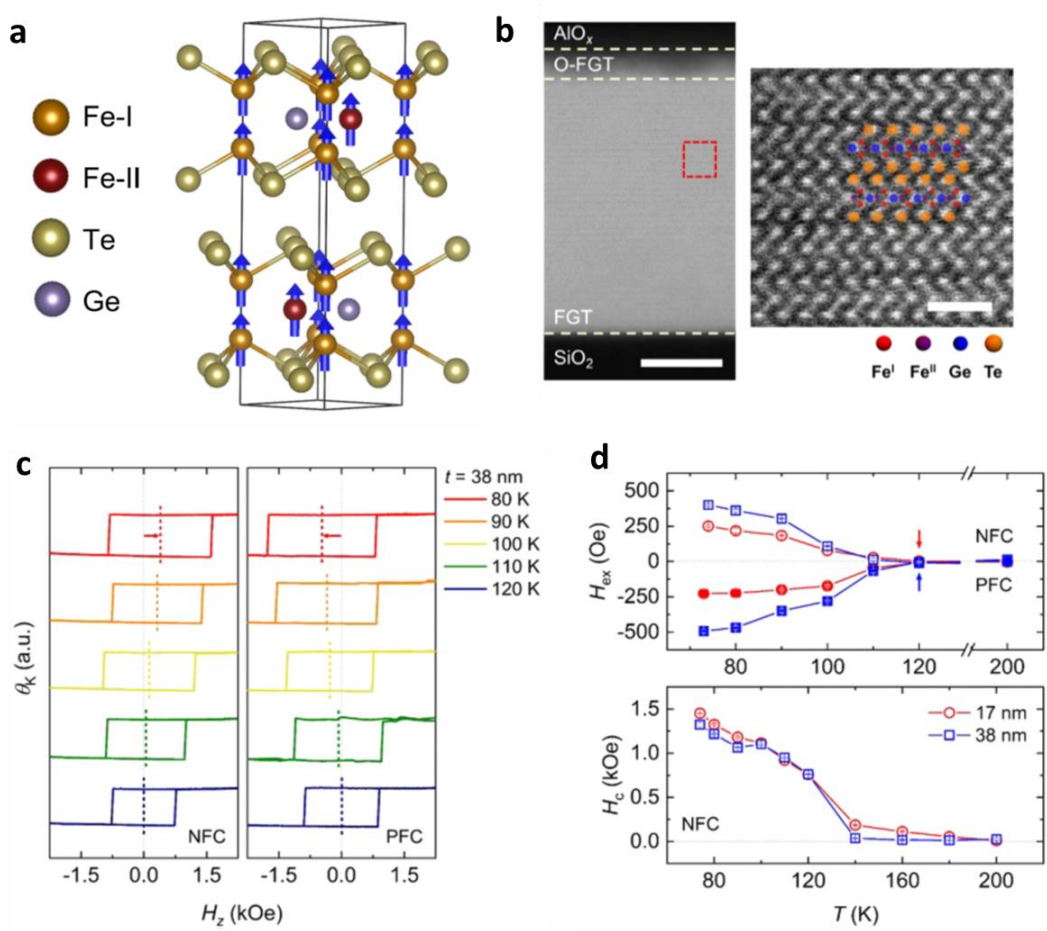

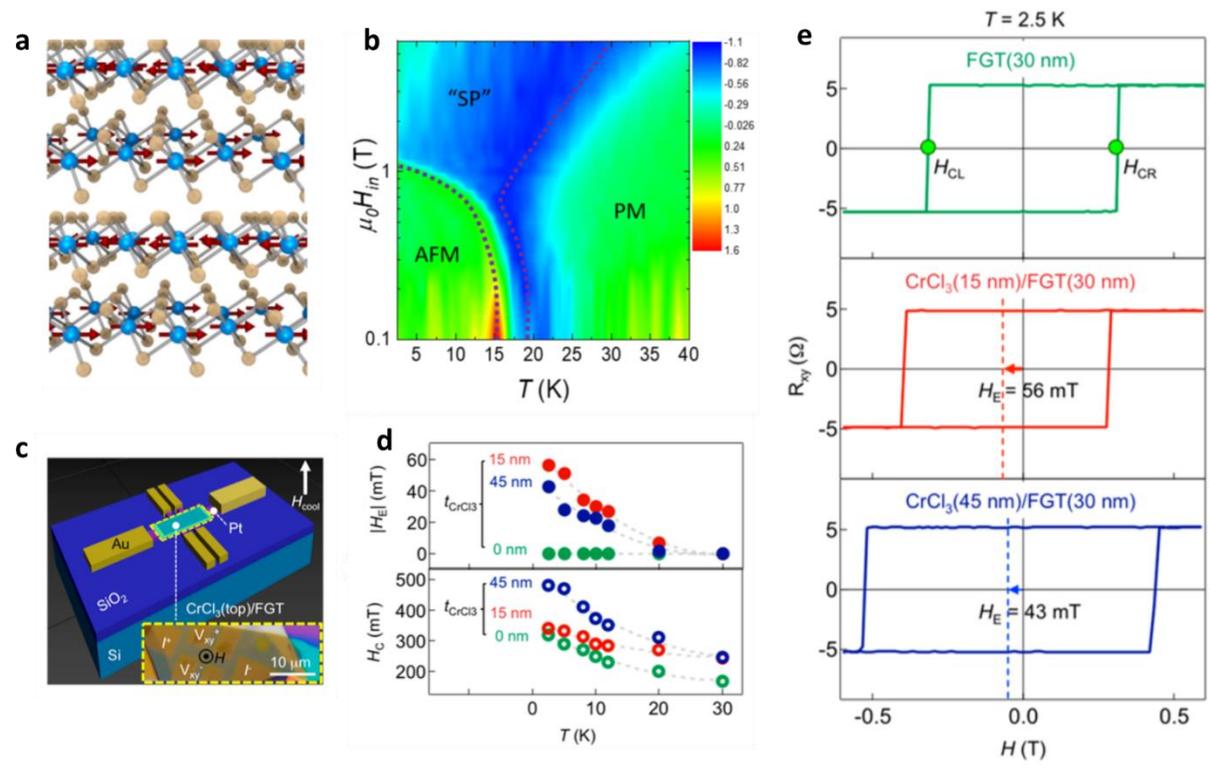

**Figure 4**

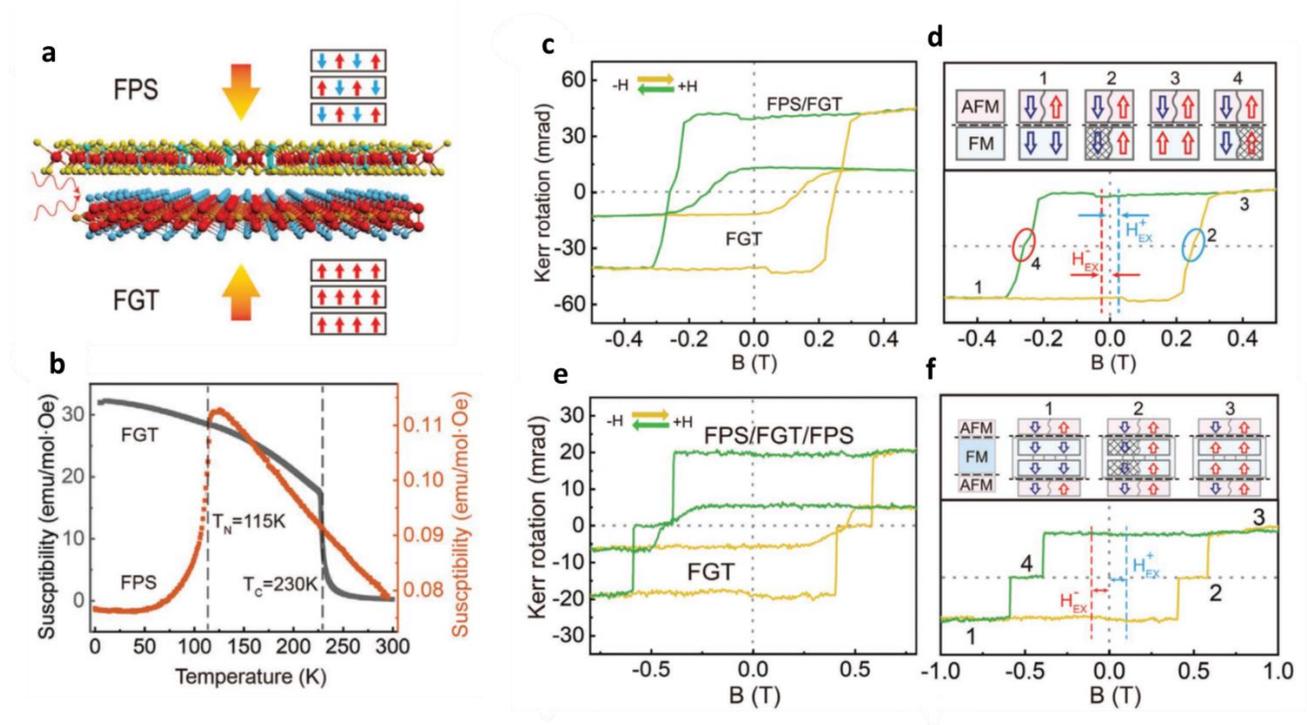

**Figure 5**

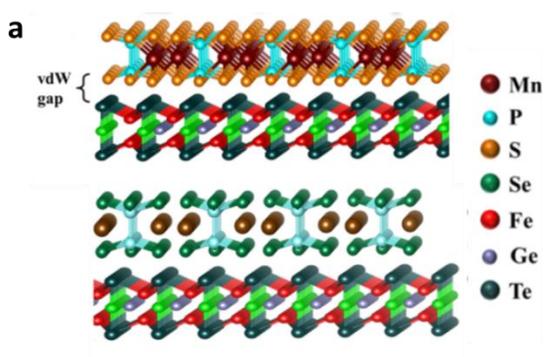

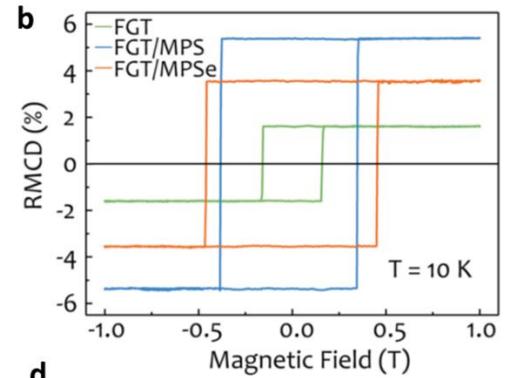

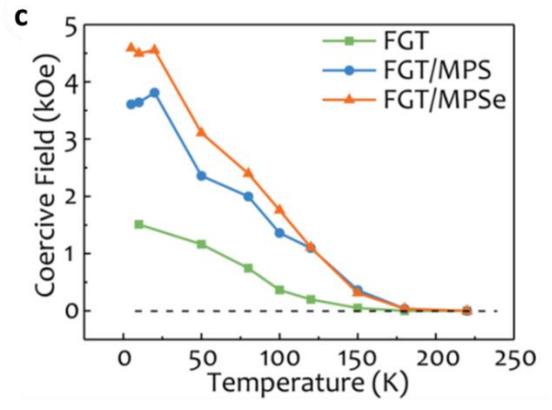

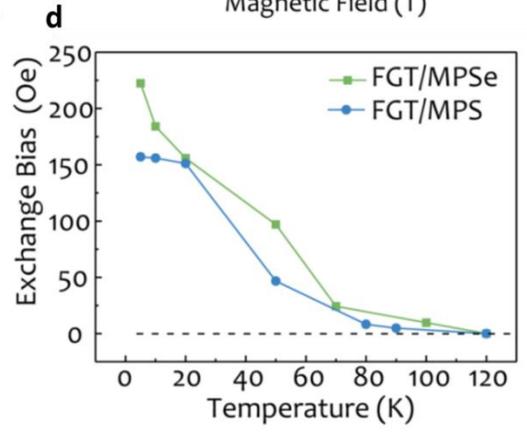

**Figure 6**

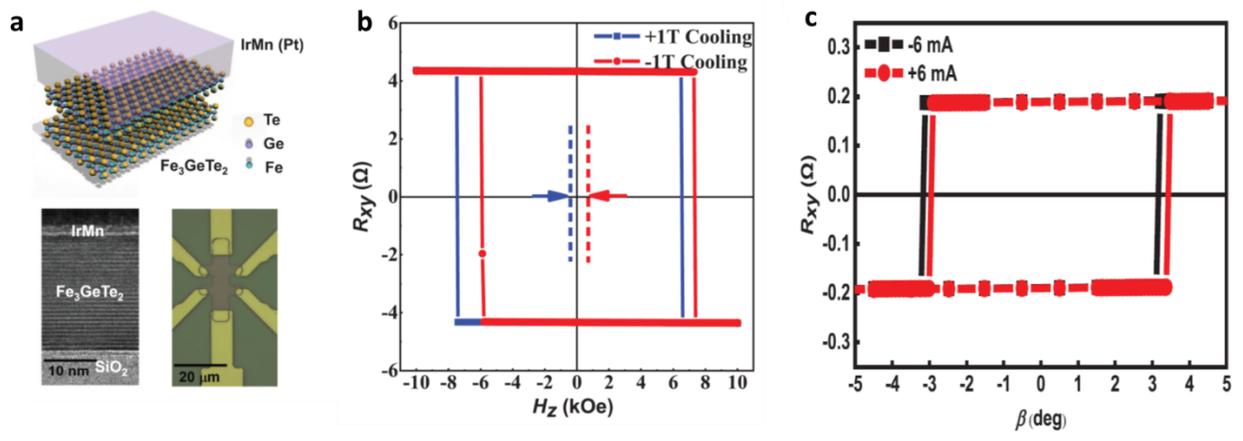

**Figure 7**

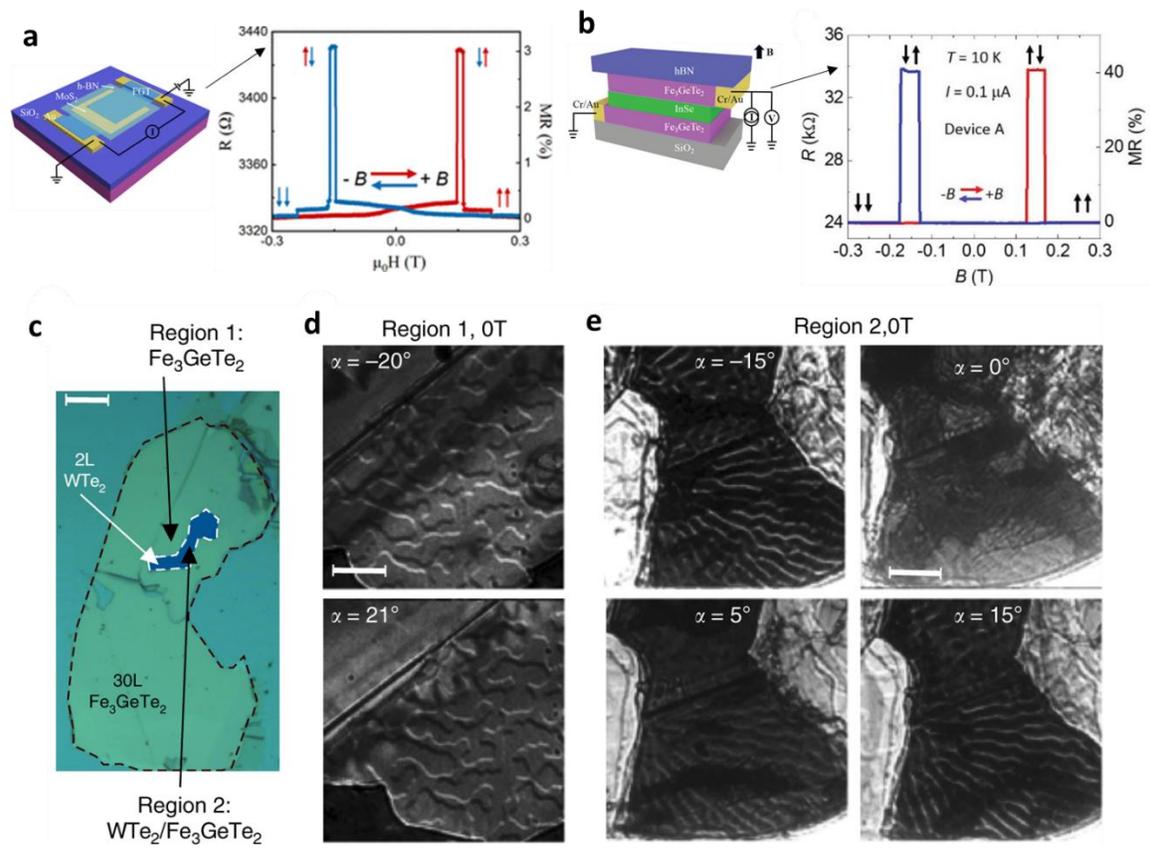

**Figure 8**

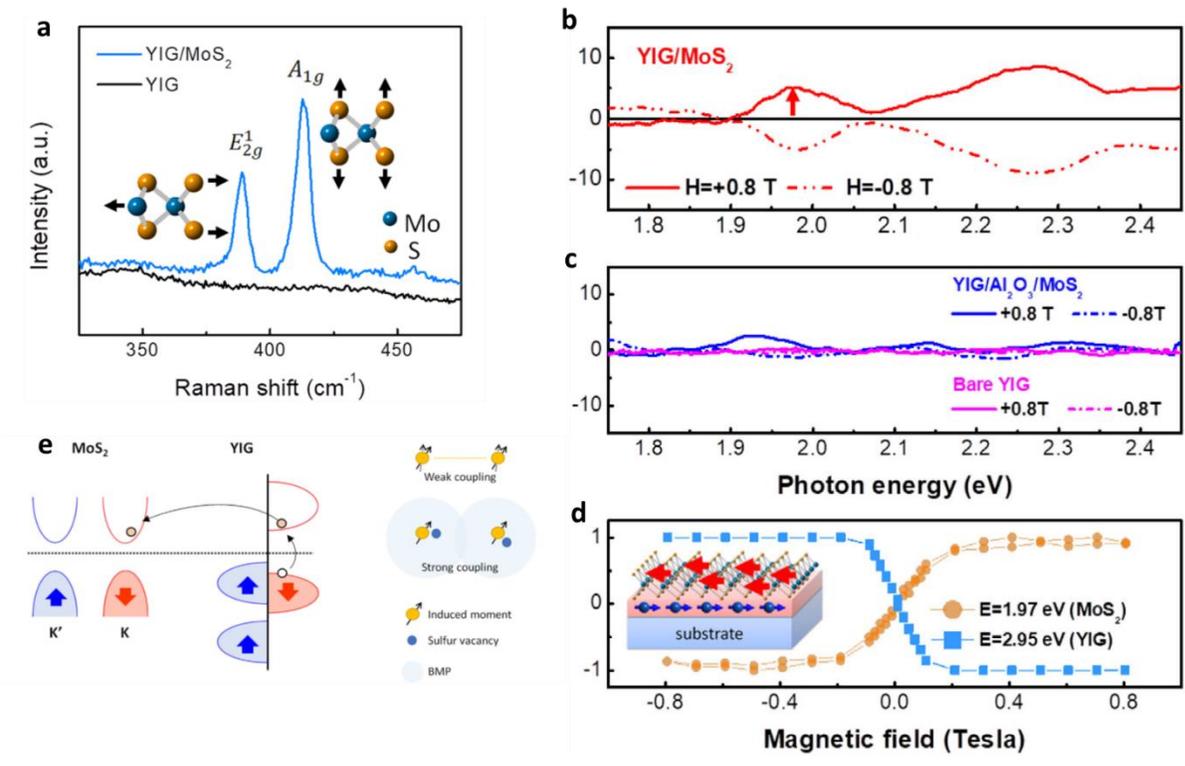

**Figure 9**

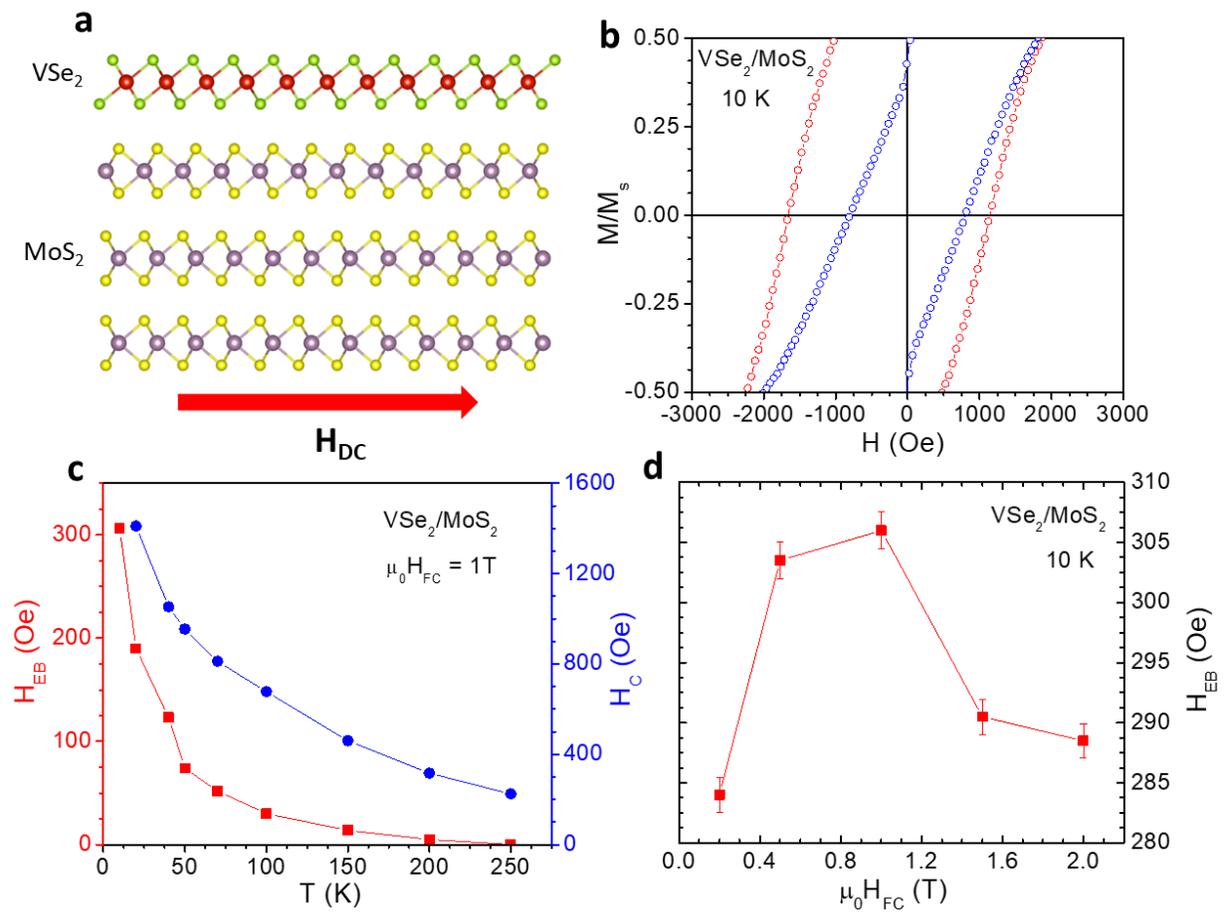

**Figure 10**

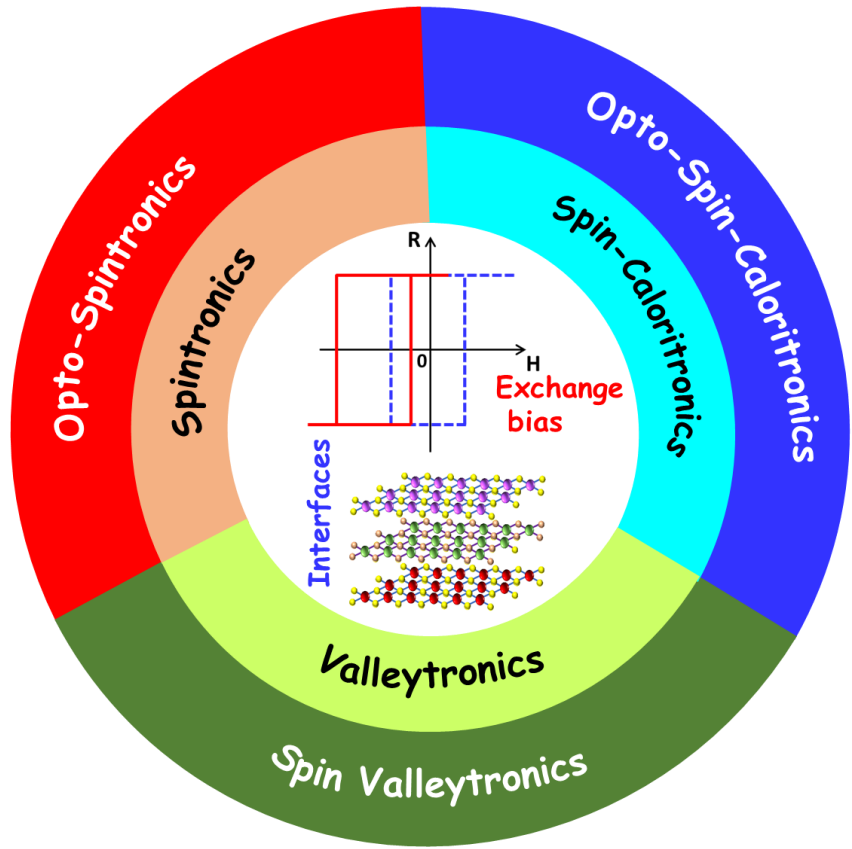